\documentclass{article}

\usepackage{times}
\usepackage{amssymb,amsmath,amsthm}
\usepackage{mathrsfs}
\usepackage{epsfig}
\usepackage{color}

\newcommand\ie{{\em i.e.}~}
\newcommand\eg{{\em e.g.}~}

\def\B{\mathscr B}
\def\C{\mathbb C}
\def\D{\mathscr D}

\def\F{\mathscr F}

\def\H{\mathcal H}

\def\K{\mathscr K}

\def\O{\mathcal O}
\def\R{\mathbb R}
\def\S{\mathscr S}

\def\U{\mathscr U}
\def\V{\mathcal V}

\def\12{{\textstyle\frac12}}
\def\<{\left\langle}
\def\>{\right\rangle}
\def\({\left(}
\def\){\right)}
\def\[{\left[}
\def\]{\right]}
\def\dom{\mathcal D}
\def\lone{\mathsf{L}^{\:\!\!1}}

\def\ltwo{\mathsf{L}^{\:\!\!2}}
\def\linf{\mathsf{L}^{\:\!\!\infty}}

\def\e{\mathop{\mathrm{e}}\nolimits}
\def\d{\mathrm{d}}

\def\im{\mathop{\mathsf{Im}}\nolimits}
\def\supp{\mathop{\mathrm{supp}}\nolimits}
\def\sgn{\mathop{\mathrm{sgn}}\nolimits}

\def\slim{\mathop{\hbox{\rm s-}\lim}\nolimits}

\newtheorem{Theorem}{Theorem}[section]
\newtheorem{Remark}[Theorem]{Remark}
\newtheorem{Lemma}[Theorem]{Lemma}
\newtheorem{Assumption}[Theorem]{Assumption}
\newtheorem{Corollary}[Theorem]{Corollary}

\newtheorem{Example}[Theorem]{Example}

\begin{document}


\title{{\Large\textbf{Time delay for dispersive systems\\
in quantum scattering theory}}}

\author{Rafael Tiedra de Aldecoa}
\date{\small
\begin{quote}
\emph{
\begin{itemize}
\item[] CNRS (UMR 8088) and Department of Mathematics, University of Cergy-Pontoise,
2 avenue Adolphe Chauvin, 95302 Cergy-Pontoise Cedex, France
\item[] \emph{E-mail:} rafael.tiedra@u-cergy.fr
\end{itemize}
}
\end{quote}
}

\maketitle


\begin{abstract}
We consider time delay and symmetrised time delay (defined in terms of sojourn times)
for quantum scattering pairs $\{H_0=h(P),H\}$, where $h(P)$ a dispersive operator of hypoelliptic-type. For instance $h(P)$ can be one of the usual elliptic operators such
as the Schr\"odinger operator $h(P)=P^2$ or the square-root Klein-Gordon operator $h(P)=\sqrt{1+P^2}$. We show under general conditions that the symmetrised time delay
exists for all smooth even localization functions. It is equal to the Eisenbud-Wigner
time delay plus a contribution due to the non-radial component of the localization
function. If the scattering operator $S$ commutes with some function of the velocity operator $\nabla h(P)$, then the time delay also exists and is equal to the
symmetrised time delay. As an illustration of our results we consider the case of a one-dimensionnal Friedrichs Hamiltonian perturbed by a finite rank potential.

Our study put into evidence an integral formula relating the operator of
differentiation with respect to the kinetic energy $h(P)$ to the time evolution of
localization operators.
\end{abstract}

\section{Introduction and main results}\label{Intro}
\setcounter{equation}{0}

One can find a large literature on the identity of Eisenbud-Wigner time delay and time
delay in quantum scattering defined in terms of sojourn times (see \cite{AC87,AS06,BGG83,dCN02,GS80,JSM72,Jen81,Mar76,Mar81,MSA92,Nak87,Nar84,Rob94,
RW89,Wan88} and references therein). However, most of the papers treat scattering
processes where the free dynamics is given by some Schr\"odinger operator. The
mathematical articles where different scattering processes are considered (such as \cite{JSM72,Mar76,Mar81,Rob94}) only furnish explicit applications in the
Schr\"odinger case. The purpose of the present paper is to fill in this gap by proving
the existence of time delay and its relation to Eisenbud-Wigner time delay for a
general class of dispersive quantum systems. Using a symmetrization argument
introduced in \cite{BO79,Mar81,Smi60} for $N$-body scattering, and rigorously applied
in \cite{AJ07,GT07,Mar75,Tie06,Tie08}, we shall treat any scattering process with free dynamics given by a regular enough pseudodifferential operator of hypoelliptic-type.

Given a real euclidean space $X$ of dimension $d\ge1$, we consider in
$\H(X):=\ltwo(X)$ the dispersive operator $H_0:=h(P)$, where $h:X\to\R$ is some
hypoelliptic function and $P\equiv(P_1,\ldots,P_d)$ is the vector momentum operator
in $\H(X)$. We also consider a selfadjoint perturbation $H$ of $H_0$ such that the
wave operators $W_\pm:=\slim\e^{itH}\e^{-itH_0}$ exist and are complete (so that the scattering operator $S:=W^*_+W_-$ is unitary). We define the usual time delay and the
symmetrised time delay for the quantum scattering system $\{H_0,H\}$ as follows. Take
a function $f\in\linf(X)$ decaying to zero sufficiently fast at infinity, and such
that $f=1$ on some bounded neighbourhood $\Sigma$ of the origin. Define for $r>0$ and
some state $\varphi\in\H(X)$ the numbers
$$
T^0_r(\varphi):=\int_\R\d t\<\e^{-itH_0}\varphi,f(Q/r)\e^{-itH_0}\varphi\>
$$
and
$$
T_r(\varphi):=\int_\R\d t\<\e^{-itH}W_-\varphi,f(Q/r)\e^{-itH}W_-\varphi\>,
$$
where $Q\equiv(Q_1,\ldots,Q_d)$ is the vector position operator in $\H(X)$. The
operator $f(Q/r)$ is approximately the projection onto the states of $\H(X)$ localized
in $r\Sigma:=\{x\in X\mid x/r\in\Sigma\}$. So, if $\varphi$ is normalised to one, $T^0_r(\varphi)$ can be roughly interpreted as the time spent by the freely evolving
state $\e^{-itH_0}\varphi$ inside the region $r\Sigma$. Similarly $T_r(\varphi)$ can
be roughly interpreted as the time spent by the associated scattering state $\e^{-itH}W_-\varphi$ inside $r\Sigma$. In consequence
$$
\tau_r^{\rm in}(\varphi):=T_r(\varphi)-T^0_r(\varphi)
$$
is approximately the time delay in $r\Sigma$ of the scattering process $\{H_0,H\}$
with incoming state $\varphi$, and
$$
\tau_r(\varphi):=T_r(\varphi)-\12\[T^0_r(\varphi)+T^0_r(S\varphi)\]
$$
is the corresponding symmetrized time delay.

In the case of the Schr\"odinger operator ($h(x)=x^2$) it is known that the existence
(and the value) of $\tau_r^{\rm in}(\varphi)$ and $\tau_r(\varphi)$ as $r\to\infty$
depend on the choice of the localization function $f$. The limit $\lim_{r\to\infty}\tau_r^{\rm in}(\varphi)$ does exist only if $f$ is radial, in which
case it is equal to Eisenbud-Wigner time delay \cite{SM92}. On another hand it has
been shown in \cite{GT07} that the limit $\lim_{r\to\infty}\tau_r(\varphi)$ does exist
for all characteristic functions $f=\chi_\Sigma$ with $\Sigma=-\Sigma$ regular enough.
In such a case the limit $\lim_{r\to\infty}\tau_r(\varphi)$ is the sum of the Eisenbud-Wigner time delay plus a term depending on the boundary $\partial\Sigma$ of $\Sigma$.

Our goal in this paper is to present a unified picture for these phenomena by
treating all scattering pairs $\{H_0\equiv h(P),H\}$, with $h$ in some natural class
of hypoelliptic functions containing $h(x)=x^2$ as a particular case (see Assumptions
\ref{h_strong}). In Section \ref{time_delay}, Theorem \ref{sym_case}, we prove under
general assumptions on $H$ and $\varphi$ the existence of the symmetrised time delay
for all smooth even functions $f$. We show that
$$
\lim_{r\to\infty}\tau_r(\varphi)=\12\<\varphi,S^*[A_f,S]\varphi\>,
$$
where $A_f$ is some explicit operator depending on $h$ and $f$ defined in Section \ref{integral}. If $f$ is radial, then $A_f$ reduces in some sense to the operator $A=-2i\,\frac\d{\d h(P)}$, and $\lim_{r\to\infty}\tau_r(\varphi)$ is equal to Eisenbud-Wigner time delay. So, if $H_0$ is purely absolutely continuous and the
scattering matrix $S(\lambda)$ is strongly continuously differentiable in the spectral representation of $H_0$, then
\begin{equation}\label{invariance}
\lim_{r\to\infty}\tau_r(\varphi)
=\int_{\sigma(H_0)}\d\lambda\,\textstyle\big\langle(\U\varphi)(\lambda),-iS(\lambda)^*
\frac{\d S(\lambda)}{\d\lambda}\,(\U\varphi)(\lambda)\big\rangle_{\H_\lambda},
\end{equation}
where $\U:\H(X)\to\int_{\sigma(H_0)}^\oplus\d\lambda\,\H_\lambda$ is a spectral transformation for $H_0$ (see Remark \ref{rem_sym} for a precise statement). If $f$ is
not radial, the limit $\lim_{r\to\infty}\tau_r(\varphi)$ is the sum of the
Eisenbud-Wigner time delay and the contribution of the non-radial component of the localization function $f$ (see Remark \ref{two_parts}). In Theorem
\ref{equal_sojourn} we show that the free sojourn times $T^0_r(\varphi)$ and $T^0_r(S\varphi)$ before and after the scattering satisfy
$$
\lim_{r\to\infty}\[T^0_r(S\varphi)-T^0_r(\varphi)\]=0
$$
if the scattering operator $S$ commutes with some appropriate function of the
velocity operator $h'(P)\equiv\nabla h(P)$. Under this circumstance the usual time
delay $\lim_{r\to\infty}\tau_r^{\rm in}(\varphi)$ also exists and is equal to
$\lim_{r\to\infty}\tau_r(\varphi)$ (see Theorem \ref{big_one}). In Corollary
\ref{two_cases} we exhibit two classes of functions $h$ for which the commutation
assumption is satisfied. Basically, these two classes of functions are the radial
functions and the polynomials of degree $1$. So, in particular, our results cover and
shed a new light on the case of the Schr\"odinger operator $h(x)=x^2$.

In Section \ref{Friedrichs}, we consider as an illustration of our approach the
simple, but instructive, case of the one-dimensionnal Friedrichs Hamiltonian
$H_0=Q$ ($H_0$ is of the form $h(P)$ after a Fourier transformation). We verify all
the assumptions of Section \ref{time_delay} when $H$ is a regular enough finite
rank perturbation of $H_0$. The main difficulty consists in showing (as in the
Schr\"odinger case \cite{ACS87,JN92}) that the scattering operator maps some dense
set into itself. Essentially this reduces to proving that the scattering matrix
$S(x)$ is sufficiently differentiable on $\R\setminus\sigma_{\rm pp}(H)$, which is
achieved by proving a stationary formula for $S(x)$ and by using higher order
commutators methods (see Lemmas \ref{rhube}-\ref{S_differentiable}). All these
results are collected in Theorem \ref{final_one}, where the formula
\begin{equation}\label{FriFri}
\lim_{r\to\infty}\tau_r^{\rm in}(\varphi)
=\lim_{r\to\infty}\tau_r(\varphi)
=-i\int_\R\d x\,|\varphi(x)|^2\overline{S(x)}S'(x)
\end{equation}
is proved for finite rank perturbations. Some comments on the relation between
Equation \eqref{FriFri} and the Birman-Krein formula are given in Remark
\ref{RemarkBirman}. The differentiability properties of the restriction operator
appearing in the expression for $S(x)$ are recalled in the appendix. 

Virtually our technics may be applied to many physical examples such as the
square-root Klein Gordon operator, the Klein-Gordon equation, the Pauli operator,
or the Dirac operator. We hope that these cases will be considered in future
publications.

Let us note that our approach relies crucially on the proof in Section \ref{integral} 
of the integral formula
\begin{equation}\label{atacama}
\lim_{r\to\infty}\int_0^\infty\d t\,\big\langle\varphi,
\big[\e^{ith(P)}f(Q/r)\e^{-ith(P)}
-\e^{-ith(P)}f(Q/r)\e^{ith(P)}\big]\varphi\big\rangle\\
=\langle\varphi,A_f\varphi\rangle.
\end{equation}
The proof of \eqref{atacama} relies in some sense on the equation
$$
\textstyle\e^{ith(P)}f\big(\frac Qr\big)\e^{-ith(P)}
=f\big(\frac{Q+th'(P)}r\big),
$$
which replaces the Alsholm-Kato formula \cite[Eq. (2.1)]{AK73}
$$
\textstyle\e^{itP^2}f\big(\frac Qr\big)\e^{-itP^2}
=\e^{-iQ^2/2t}f\big(\frac{tP}r\big)\e^{iQ^2/2t}
$$
of the Schr\"odinger case. We think that Formula \eqref{atacama} is interesting on
its own, since it relates (when $f$ is radial) the time evolution of the localization operator $f(Q/r)$ to the operator of differentiation with respect to the kinetic
energy $h(P)$.

As a final comment, we would like to emphasize that this paper shows that the Eisenbud-Wigner operator $-iS(\lambda)^*\frac{\d S(\lambda)}{\d\lambda}$ is the
on-shell value of a time delay operator (symmetrised or not), not only for Schr\"odinger-type scattering systems, but for a large class of scattering pairs
$\{H_0,H\}$. This was not so clear from the very beginning.

We finally mention the papers \cite{Pus08,Tam08} for recent works on time delay.

\section{Averaged localization functions}\label{averaged}
\setcounter{equation}{0}

In this section we collect results on a class of averaged localization functions which appears naturally when dealing with quantum time delay. We start by fixing some
notations which will be freely used throughout the paper.

We write $|\cdot|$ for the norm in $X$, set $\<\cdot\>:=(1+|\cdot|^2)^{1/2}$, and use $\underline\d x:=(2\pi)^{-d/2}\d x$ as measure on $X$ ($\d x$ is the usual euclidean
measure on $X$). We denote by $x\cdot y$ the scalar product of $x,y\in X$. Sometimes
we identify $X$ with $\R^d$ by choosing in $X$ an orthonormal basis
$\V:=\{v_1,\ldots,v_d\}$. Given a function $g\in C^1(X;\C)$, we write $g'(x)$ for the derivative of $g$ at $x$, \ie $g(x+h)=g(x)+h\cdot g'(x)+o(|h|)$ for $h\in X$ with
$|h|$ sufficiently small. For higher order derivatives, we use the multi-index
notation. A multi-index $\alpha$ is a $d$-tuple $(\alpha_1,\ldots,\alpha_d)$ of
integers $\alpha_j\ge0$ such that
$$
|\alpha|:=\alpha_1+\ldots+\alpha_d,
\qquad\alpha!:=\alpha_1\cdots\alpha_d,
\qquad\partial^\alpha:=\partial_1^{\alpha_1}\ldots\partial_d^{\alpha_d},
$$
and
$$
x^\alpha:=x_1^{\alpha_1}\cdots x_d^{\alpha_d}\quad{\rm if}\quad
x=x_1v_1+\cdots x_dv_d\in X\quad(x_j\in\R).
$$
The Hilbert space $\H(X)=\ltwo(X)$ is endowed with its usual norm $\|\cdot\|$ and
scalar product $\<\cdot,\cdot\>$. The $j$-th components of $P$ and $Q$ with respect
to $\V$ act as $(P_j\varphi)(x):=-i(\partial_j\varphi)(x)$ and
$(Q_j\varphi)(x):=x_j\varphi(x)$ in $\H(X)$.

\begin{Assumption}\label{assumption_f}
The function $f\in\linf(X)$ satisfies the following conditions:
\begin{enumerate}
\item[(i)] There exists $\rho>0$ such that
$|f(x)|\le{\rm Const.}\,\langle x\rangle^{-\rho}$ for a.e. $x\in X$.
\item[(ii)] $f=1$ on a bounded neighbourhood of~~$0$.
\end{enumerate}
\end{Assumption}

It is clear that $\slim_{r\to\infty}f(Q/r)=1$ if $f$ satisfies Assumption \ref{assumption_f}. Furthermore, one has for each $x\in X\setminus\{0\}$
$$
\left|\int_0^\infty\frac{\d\mu}\mu\[f(\mu x)-\chi_{[0,1]}(\mu)\]\right|
\le\int_0^1\frac{\d\mu}\mu\,|f(\mu x)-1|
+{\rm Const.}\int_1^{+\infty}\d\mu\,\mu^{-(1+\rho)}
<\infty.
$$
Therefore the function $R_f:X\setminus\{0\}\to\C$ given by
$$
R_f(x):=\int_0^{+\infty}\frac{\d\mu}\mu\[f(\mu x)-\chi_{[0,1]}(\mu)\]
$$
is well-defined (see \cite[Sec. 2]{GT07} and \cite[Sec. 2]{Tie08} for a similar
definition).

In the next lemma we establish some differentiability properties of $R_f$. The symbol $\S(X)$ stands for the Schwartz space on $X$.

\begin{Lemma}\label{function_R}
Let $f$ satisfy Assumption \ref{assumption_f}. Then
\begin{enumerate}
\item[(a)] For all $j\in\{1,2,\ldots,d\}$ and $x\in X$, assume that
$(\partial_jf)(x)$ exists and satisfies
$|(\partial_jf)(x)|\le{\rm Const.}\<x\>^{-(1+\rho)}$. Then $R_f$ is differentiable
on $X\setminus\{0\}$, and its derivative is given by
\begin{equation}\label{diff_R_f}
R_f'(x)=\int_0^\infty\d\mu\,f'(\mu x).
\end{equation}
Moreover, $R_f$ belongs to $C^\infty(X\setminus\{0\})$ if $f\in\S(X)$.
\item[(b)] Assume that $R_f$ belongs to $C^m(X\setminus\{0\})$ for some $m\ge1$. Then
one has for each $x\in X\setminus\{0\}$ and $t>0$ the homogeneity properties
\begin{align}
x\cdot R_f'(x)&=-1,\label{minusone}\\
t^{|\alpha|}(\partial^\alpha R_f)(tx)
&=(\partial^\alpha R_f)(x),\label{R_f_alpha}
\end{align}
where $\alpha$ is a multi-index with $1\le|\alpha|\le m$.
\item[(c)] Assume that $f$ is radial, \ie there exists $f_0\in\linf(\R)$ such that $f(x)=f_0(|x|)$ for a.e. $x\in X$. Then $R_f$ belongs to $C^\infty(X\setminus\{0\})$,
and $R_f'(x)=-x^{-2}x$.
\end{enumerate}
\end{Lemma}

\begin{proof}
(a) The claim is a consequence of standard results on differentation under the integral
(see \eg \cite[Chap. 13, Lemma 2.2]{Lan83}).

(b) Let $x\in X\setminus\{0\}$ and $t>0$. Then one has
\begin{align}
R_f(tx)&=\int_0^\infty\frac{\d\mu}\mu\[f(\mu tx)-\chi_{[0,1]}(\mu)\]\nonumber\\
&=\int_0^\infty\frac{\d\mu}\mu\[f(\mu)-\chi_{[0,1]}(\mu)\]
+\int_0^\infty\frac{\d\mu}\mu\[\chi_{[0,1]}(\mu)-\chi_{[0,t]}(\mu)\]\nonumber\\
&=R_f(x)-\ln t,\label{logarithm}
\end{align}
and the claim follows by taking derivatives with respect to $t$ and $x$.

(c) For $x\in X\setminus\{0\}$, one gets
$$
R_{f_0}(1)=R_f(x)+\ln|x|,
$$
by putting $t=|x|^{-1}$ in Equation \eqref{logarithm}. This implies the claim.
\end{proof}

In the sequel we shall also need the function $F_f:X\setminus\{0\}\to\C$ defined by
$$
F_f(x):=\int_\R\d\mu\,f(\mu x).
$$
The function $F_f$ satisfies several properties as $R_f$. Here we only note that
$F_f$ is well-defined if $f$ satisfies Assumption \ref{assumption_f}.(i) with
$\rho>1$, and that
\begin{equation}\label{homo_f}
F_f(x)=tF_f(tx)
\quad\textrm{for each }t>0\textrm{ and each }x\in X\setminus\{0\}.
\end{equation}
Physically, if $p\in\R^d$ and $f\ge0$, then the number
$F_f(p)\equiv\int_\R\d t\,f(tp)$ can be seen as the sojourn time in the region defined
by the localization function $f$ of a free classical particle moving along the
trajectory $\R\ni t\mapsto x(t):=tp$.

\section{Integral formula for $\boldsymbol{H_0=h(P)}$}\label{integral}

Given a function $h\in C^1(X;\R)$, we denote by $\kappa(h)$ the set of critical values
of $h$, \ie
$$
\kappa(h):=\{\lambda\in\R\mid\exists x\in X\textrm{ such that }h(x)=\lambda
\textrm{ and }h'(x)=0\}.
$$
The size and the topology of $\kappa(h)$ depends on the regularity and the behaviour
of the function $h$. Here we only recall some properties of $\kappa(h)$ (see
\cite[Sec. 7.6.2]{ABG} for more details): 
\begin{enumerate}
\item $H_0=h(P)$, whose spectrum is $\sigma(H_0)=\overline{h(X)}$, has purely
absolutely continuous spectrum in $\sigma(H_0)\setminus\kappa(h)$.
\item $H_0$ is purely absolutely continuous if $h^{-1}(\kappa(h))$ has measure zero.
\item $\kappa(h)$ has measure zero if $h\in C^d(X;\R)$, with $d\ge1$ the dimension
of $X$.
\item $\kappa(h)$ is finite if $h$ is a polynomial.
\item $\kappa(h)$ is closed if $|h(x)|+|h'(x)|\to\infty$ as $|x|\to\infty$.
\end{enumerate}

\begin{Assumption}\label{h}
The function $h:X\to\R$ is of class $C^m$ for some $m\ge2$, and satisfies the
following conditions:
\begin{enumerate}
\item[(i)] $|h(x)|\to\infty$~~as~~$|x|\to\infty$.
\item[(ii)] $\sum_{|\alpha|\le m}|(\partial^\alpha h)(x)|\le{\rm Const.}\;\!(1+|h(x)|)$.
\end{enumerate}
\end{Assumption}

For each $s,t\in\R$, we denote by $\H^s_t(X)$ the usual weighted Sobolev space over
$X$, namely the completion of $\S(X)$ for the norm
$\|\varphi\|_{\H^s_t(X)}:=\|\langle P\rangle^s\langle Q\rangle^t\varphi\|$. We also
set $\H^s(X):=\H^s_0(X)$ and $\H_t(X):=\H_t^0(X)$, and for each $t\ge0$ we define
$$
\D_t^0(X):=\big\{\varphi\in\H_t(X)\mid\eta(h(P))\varphi=\varphi
\textrm{ for some }\eta\in C^\infty_{\rm c}\big(\R\setminus\kappa(h)\big)\big\}.
$$
The set $\D_t^0(X)$ is included in the subspace $\H_{\rm ac}(H_0)$ of absolute
continuity of $H_0$, $\D_t^0(X)$ is dense in $\H(X)$ if $h^{-1}(\kappa(h))$ has measure zero, and $\D_{t_1}^0(X)\subset\D_{t_2}^0(X)$ if $t_1\ge t_2$.

\begin{Lemma}\label{lemma_A_f}
Let $f$ satisfy Assumption \ref{assumption_f}, assume that $R_f$ belongs to $C^2(X\setminus\{0\})$, and let $h$ satisfy Assumption \ref{h}. Then the operator given
by the formal expression
\begin{equation}\label{A_f}
A_f:=Q\cdot R_f'(h'(P))+R_f'(h'(P))\cdot Q
\end{equation}
is well-defined on $\D_1^0(X)$. In particular $\{A_f,\D_1^0(X)\}$ is symmetric if $f$
is real and $h^{-1}(\kappa(h))$ has measure zero.
\end{Lemma}

\begin{proof}
Let $\varphi\in\D_1^0(X)$ and choose
$\eta\in C^\infty_{\rm c}\big(\R\setminus\kappa(h)\big)$ such that $\eta(h(P))\varphi=\varphi$. Then there exists $\textsc c>0$ such that
$|h'(x)|>\textsc c$ for all $x\in h^{-1}(\supp\eta)$, due to Assumption \ref{h}.(i)
(see the discussion after \cite[Prop. 7.6.6]{ABG} for details). This together with Assumption \ref{h}.(ii) implies that
\begin{equation}\label{bo-bound}
\big\||h'(P)|^{-2}\eta(h(P))(\partial^\alpha h)(P)\big\|<\infty
\quad{\rm and}\quad\big\||h'(P)|^{-1}\eta(h(P))\big\|<\infty
\end{equation}
for any multi-index $\alpha$ with $|\alpha|\le2$. Furthermore the operator
$(\partial^\alpha R_f)\big(\frac{h'(P)}{|h'(P)|}\big)$ is also bounded for $\alpha$
with $|\alpha|\le2$, due to the compacity of $(\partial^\alpha R_f)(\mathbb S^{d-1})$.
Therefore, using Formula \eqref{R_f_alpha} with $t=|x|^{-1}$, we get the estimate
\begin{align*}
\|A_f\varphi\|
&=\textstyle\big\|\big\{i\sum_{j\le d}(\partial_jh)'(P)
\cdot(\partial_jR_f)'(h'(P))+2R_f'(h'(P))\cdot Q\big\}\eta(h(P))\varphi\big\|\\
&\le\textstyle\Big\|\sum_{j\le d}|h'(P)|^{-2}\eta(h(P))(\partial_jh)'(P)
\cdot(\partial_jR_f)'\big(\frac{h'(P)}{|h'(P)|}\big)\varphi\Big\|\\
&\quad\textstyle+{\rm Const.}\,\big\|\big\{|h'(P)|^{-1}\eta(h(P))
R_f'\big(\frac{h'(P)}{|h'(P)|}\big)\cdot Q\big\}\varphi\big\|\\
&\le{\rm Const.}\,\|\langle Q\rangle\varphi\|,
\end{align*}
which implies the claim.
\end{proof}

There are at least two cases where the operator $A_f$ takes a simple form. First,
suppose that $h$ is a polynomial of degree $1$, \ie $h(x)=v_0+v\cdot x$ for some
$v_0\in\R$, $v\in X\setminus\{0\}$. Then the operator $R_f'(h'(P))$ reduces to the
constant vector $R_f'(v)$, and
$$
A_f:=2R_f'(v)\cdot Q.
$$
Second, suppose that $f$ is radial. Then one has $R_f'(x)=-x^{-2}x$ due to Lemma \ref{function_R}.(c), and $A_f$ reduces to the operator
\begin{equation}\label{A}
A:=-\textstyle\big(Q\cdot\frac{h'(P)}{h'(P)^2}+\frac{h'(P)}{h'(P)^2}\cdot Q\big).
\end{equation}
For instance, in the particular case where $h(x)=h_0(|x|)$ with $h_0'\ge0$, one gets
\begin{equation}\label{A_0}
A_0:=-\textstyle\big(Q\cdot\frac P{|P|h_0'(|P|)}+\frac P{|P|h_0'(|P|)}\cdot Q\big).
\end{equation}

Next Theorem is somehow related to the usual result on the asymptotic velocity for Hamiltonians $H_0=h(P)$ (see \eg \cite{JLN}, \cite[Sec. 2]{Rob80},
\cite[Thm. 7.1.29]{Hor83}, and \cite[Sec. 7.C]{ABG}). The symbol $\F$ stands for the
Fourier transformation.

\begin{Theorem}\label{for_Schwartz}
Let $f\in\S(X)$ be an even function such that $f=1$ on a bounded neighbourhood of $0$.
Let $h$ satisfy Assumption \ref{h} with $m\ge3$. Then we have for each
$\varphi\in\D_2^0(X)$
\begin{equation}\label{bisoufe}
\lim_{r\to\infty}\int_0^\infty\d t\,\big\langle\varphi,
\big[\e^{ith(P)}f(Q/r)\e^{-ith(P)}
-\e^{-ith(P)}f(Q/r)\e^{ith(P)}\big]\varphi\big\rangle\\
=\langle\varphi,A_f\varphi\rangle.
\end{equation}
\end{Theorem}

\begin{proof}
(i) Let $\varphi\in\D_2^0(X)$, take a real
$\eta\in C^\infty_{\rm c}\big(\R\setminus\kappa(h)\big)$ such that
$\eta(h(P))\varphi=\varphi$, and set $\eta_t(P):=\e^{ith(P)}\eta(h(P))$. Then we have
\begin{align}
&\big\langle\varphi,\big[\e^{ith(P)}f(Q/r)\e^{-ith(P)}
-\e^{-ith(P)}f(Q/r)\e^{ith(P)}\big]\varphi\big\rangle\nonumber\\
&=\int_X\underline\d x\,(\F f)(x)
\big\langle\varphi,\big[\eta_t(P)\e^{i\frac xr\cdot Q}\eta_{-t}(P)
-\eta_{-t}(P)\e^{i\frac xr\cdot Q}\eta_t(P)\big]\varphi\big\rangle\nonumber\\
&=\int_X\underline\d x\,(\F f)(x)\big\langle\varphi,
\big[\e^{i\frac xr\cdot Q}\eta_t\big(P+\textstyle\frac xr\big)\eta_{-t}(P)
-\eta_{-t}(P)\eta_t\big(P-\textstyle\frac xr\big)\e^{i\frac xr\cdot Q}\big]
\varphi\big\rangle\nonumber\\
&=\int_X\underline\d x\,(\F f)(x)\big\langle\varphi,
\big\{\big(\e^{i\frac xr\cdot Q}-1\big)\eta_t\big(P+\textstyle\frac xr\big)
\eta_{-t}(P)\label{pave}\\
&\qquad\qquad+\eta_{-t}(P)\big[\eta_t\big(P+\textstyle\frac xr\big)
-\eta_t\big(P-\textstyle\frac xr\big)\big]
-\eta_{-t}(P)\eta_t\big(P-\textstyle\frac xr\big)
\big(\e^{i\frac xr\cdot Q}-1\big)\big\}\varphi\big\rangle.\nonumber
\end{align}
Since $f$ is even, $\F f$ is also even, and
$$
\int_X\underline\d x\,(\F f)(x)\<\varphi,
\eta_{-t}(P)\big[\eta_t\big(P+\textstyle\frac xr\big)
-\eta_t\big(P-\textstyle\frac xr\big)\big]\varphi\>=0.
$$
Thus Formula \eqref{pave} and the change of variables $\mu:=t/r$, $\nu:=1/r$, give
\begin{align}
&\lim_{r\to\infty}\int_0^\infty\d t\,\big\langle\varphi,\big[\e^{ith(P)}f(Q/r)\e^{-ith(P)}
-\e^{-ith(P)}f(Q/r)\e^{ith(P)}\big]\varphi\big\rangle\nonumber\\
&=\lim_{\nu\searrow0}\int_0^\infty\d\mu\int_X\underline\d x\,K(\nu,\mu,x),\label{limit}
\end{align}
where
\begin{align*}
K(\nu,\mu,x):=(\F f)(x)\big\langle\varphi,
\big\{&\textstyle\frac1\nu\big(\e^{i\nu x\cdot Q}-1\big)
\eta(h(P+\nu x))\e^{i\frac\mu\nu[h(P+\nu x)-h(P)]}\nonumber\\
&-\eta(h(P-\nu x))\e^{i\frac\mu\nu[h(P-\nu x)-h(P)]}
\textstyle\frac1\nu\big(\e^{i\nu x\cdot Q}-1\big)\big\}\varphi\big\rangle.
\end{align*}

(ii) To prove the statement, we shall show that one may interchange the limit and the integrals in \eqref{limit}, by invoking Lebesgue's dominated convergence theorem. This
will be done in (iii) below. If one assumes that these interchanges are justified for
the moment, then direct calculations using the parity of $f$, Lemma
\ref{function_R}.(a), and Lemma \ref{lemma_A_f} give
\begin{align*}
&\lim_{r\to\infty}\int_0^\infty\d t\,\big\langle\varphi,\big[\e^{ith(P)}f(Q/r)\e^{-ith(P)}
-\e^{-ith(P)}f(Q/r)\e^{ith(P)}\big]\varphi\big\rangle\\
&=i\int_0^\infty\d\mu\,\int_X\underline\d x\,(\F f)(x)
\big\{\big\langle\(x\cdot Q\)\varphi,\e^{i\mu x\cdot h'(P)},\varphi\big\rangle
-\big\langle\varphi,\e^{-i\mu x\cdot h'(P)}\(x\cdot Q\)\varphi\big\rangle\big\}\\
&=\sum_{j\le d}\int_0^\infty\d\mu\,\int_X\underline\d x\,[\F(\partial_jf)](x)
\big[\big\langle Q_j\varphi,\e^{i\mu x\cdot h'(P)},\varphi\big\rangle
+\big\langle\varphi,\e^{i\mu x\cdot h'(P)}Q_j\varphi\big\rangle\big]\\
&=\sum_{j\le d}\int_0^\infty\d\mu\,
\big[\big\langle Q_j\varphi,(\partial_jf)(\mu h'(P))\varphi\big\rangle
+\big\langle\varphi,(\partial_jf)(\mu h'(P))Q_j\varphi\big\rangle\big]\\
&=\langle\varphi,A_f\varphi\rangle.
\end{align*}

(iii) To interchange the limit $\nu\searrow0$ and the integration over $\mu$ in
Equation \eqref{limit}, one has to bound $\int_X\underline\d x\,K(\nu,\mu,x)$ uniformly
in $\nu$ by a function in $\lone((0,\infty),\d\mu)$. We begin with the first term
of $\int_X\underline\d x\,K(\nu,\mu,x)$:
\begin{align*}
&K_1(\nu,\mu)\\
&:=\int_X\underline\d x\,(\F f)(x)
\<\langle Q\rangle^2\varphi,
\textstyle\frac1\nu\big(\e^{i\nu x\cdot Q}-1\big)
\langle Q\rangle^{-2}\eta(h(P+\nu x))
\e^{i\frac\mu\nu[h(P+\nu x)-h(P)]}\varphi\>.
\end{align*}
One has
\begin{equation}\label{nu_and_x}
\big\|\textstyle\frac1\nu\big(\e^{i\nu x\cdot Q}-1\big)\<Q\>^{-2}\big\|
\le{\rm Const.}\,|x|
\end{equation}
due to the spectral theorem and the mean value theorem. Since $\F f\in\S(X)$ it follows
that
\begin{equation}\label{mu_small}
\big|K_1(\nu,\mu)\big|\le{\rm Const.},
\end{equation}
and thus $K_1(\nu,\mu)$ is bounded uniformly in $\nu$ by a function in
$\lone((0,1],\d\mu)$.

For the case $\mu>1$ we recall that there exists $\textsc c>0$ such that
$|h'(x)|>\textsc c$ for all $x\in h^{-1}(\supp\eta)$, due to Assumption \ref{h}.(i). Therefore the operator
$$
A_{j,\nu}(x):=(\F f)(x)\textstyle\frac1\nu\big(\e^{i\nu x\cdot Q}-1\big)\<Q\>^{-2}
\textstyle\frac{\eta(h(P+\nu x))(\partial_jh)(P+\nu x)}{|h'(P+\nu x)|^2}
$$
satisfies for any integer $k\ge1$ the bound
$$
\|A_{j,\nu}(x)\|\le{\rm Const.}\<x\>^{-k},
$$
due to Equations \eqref{bo-bound}, \eqref{nu_and_x}, and the rapid decay of $\F f$.
So $K_1(\nu,\mu)$ can be written as
$$
K_1(\nu,\mu)=-i\mu^{-1}\sum_{j\le d}\int_X\underline\d x
\<\langle Q\rangle^2\varphi,A_{j,\nu}(x)(\partial_jB_{\nu,\mu})(x)\varphi\>,
$$
with $B_{\nu,\mu}(x):=\e^{i\frac\mu\nu[h(P+\nu x)-h(P)]}$. Moreover lengthy, but
direct, calculations using Equation \eqref{nu_and_x} and Assumption \ref{h}.(ii)
show that
$$
\|(\partial_jA_{j,\nu})(x)\|\le{\rm Const.}\,(1+|\nu|)\<x\>^{-k}
$$
and
\begin{equation}\label{chmolo}
\Big\|\partial_\ell(\partial_jA_{j,\nu})(x)
\textstyle\frac{(\partial_\ell h)(P+\nu x)}{|h'(P+\nu x)|^2}\Big\|
\le{\rm Const.}\(1+|\nu|+\nu^2\)\<x\>^{-k}
\end{equation}
for any integer $k\ge1$. Therefore one can perform two successive integrations by parts (with vanishing boundary contributions) and obtain
\begin{align*}
K_1(\nu,\mu)&=i\mu^{-1}\sum_{j\le d}\int_X\underline\d x\<\langle Q\rangle^2\varphi
,(\partial_jA_\nu)(x)B_{\nu,\mu}(x)\varphi\>\\
&=-\mu^{-2}\sum_{j,\ell\le d}\int_X\underline\d x\,\big\langle\langle Q\rangle^2\varphi
,\big\{\partial_\ell(\partial_jA_{j,\nu})(x)
\textstyle\frac{(\partial_\ell h)(P+\nu x)}{|h'(P+\nu x)|^2}\big\}
B_{\nu,\mu}(x)\varphi\big\rangle.\label{two_by_parts}
\end{align*}
This together with Formula \eqref{chmolo} implies that
\begin{equation}\label{mu_big}
\big|K_1(\nu,\mu)\big|\le{\rm Const.}\,\mu^{-2}
\quad\textrm{for each }\nu<1\textrm{ and each }\mu>1.
\end{equation}
The combination of the bounds \eqref{mu_small} and \eqref{mu_big} shows that
$K_1(\nu,\mu)$ is bounded uniformly for $\nu<1$ by a function in
$\lone((0,\infty),\d\mu)$. Since similar arguments shows that the same holds for the
second term of $\int_X\underline\d x\,K(\nu,\mu,x)$, one can interchange the limit
$\nu\searrow0$ and the integration over $\mu$ in Equation \eqref{limit}.

The interchange of the limit $\nu\searrow0$ and the integration over $x$
in \eqref{limit} is justified by the bound
$$
\big|K(\nu,\mu,x)\big|\le{\rm Const.}\,\big|x(\F f)(x)\big|,
$$
which follows from Formula \eqref{nu_and_x}.
\end{proof}

\begin{Remark}\label{pipo}
We strongly believe that Formula \eqref{bisoufe} remains true for a large class
of non-smooth even localization functions $f$ (such as characteristic functions, for instance). In the particular cases of the Schr\"odinger operator $h(x)=x^2$
and the one-dimensional Friedrichs model $h(x)=x$, similar results suggest that
$f$ only has to decay to $0$ sufficiently fast at infinity (see \cite[Prop. 4.5]{GT07}
and Section \ref{Preliminaries}). Unfortunately, in the general situation, we have not
been able to extend the proof of Theorem \ref{for_Schwartz} to such a class of functions. Only minor ameliorations, not worth to mention, have been obtained. 
\end{Remark}

Next result follows directly from Lemma \ref{function_R}.(c) and Theorem
\ref{for_Schwartz}.

\begin{Corollary}\label{radial_case}
Let $f\in\S(X)$ be a radial function such that $f=1$ on a bounded neighbourhood of $0$.
Let $h$ satisfy Assumption \ref{h} with $m\ge3$. Then we have for each
$\varphi\in\D_2^0(X)$
\begin{equation}\label{energy}
\lim_{r\to\infty}\int_0^\infty\d t\,
\big\langle\varphi,\big[\e^{ith(P)}f(Q/r)\e^{-ith(P)}
-\e^{-ith(P)}f(Q/r)\e^{ith(P)}\big]\varphi\big\rangle\\
=\langle\varphi,A\varphi\rangle,
\end{equation}
with $A$ defined by \eqref{A}.
\end{Corollary}

The rest of the section is devoted to the interpretation of Formula
\eqref{energy}. We consider first the operator $A$ on the r.h.s. One has
for each $\varphi\in\D^0_1(X)$
\begin{equation}\label{santiago}
[A,h(P)]\varphi=-2i\varphi,
\end{equation}
which suggest that $A=-2i\,\frac\d{\d h(P)}$, with a slight abuse of notation. Thus,
formally, $\frac i2A$ can be seen as the operator of differentiation with respect to the kinetic energy $h(P)$. In fact, this affirmation could be turned into a rigorous statement in many concrete situations. As an example, we present two particular cases where rigorous formulas can be easily obtained.

{\bf Case 1:} Suppose that $h$ is a polynomial of degree $1$ satisfying the hypotheses
of Corollary \ref{radial_case}. Then $h(x)=v_0+v\cdot x$ for some $v_0\in\R$,
$v\in X\setminus\{0\}$, and we have $h(X)=\R$ and $\kappa(h)=\varnothing$. So $H_0$ has purely absolutey continuous spectrum $\sigma(H_0)=\sigma_{\rm ac}(H_0)=\R$. Moreover the operators $A\equiv-2\frac v{v^2}\cdot Q$ and $h(P)\equiv v_0+v\cdot P$ are selfajoint,
and have $\S(X)$ as a common core. The associated unitary groups $U(t):=\e^{itA}$ and $V(s):=\e^{ish(P)}$ are continuous, and satisfy the Weyl relations
$$
U(t/2)V(s)=\e^{its}V(s)U(t/2).
$$
It follows by the Stone-von Neumann theorem \cite[VIII.14]{RSI} that there exists
a unitary operator $\U_1:\H(X)\to\ltwo(\R;\C^N,\d\lambda)$, with $N$ finite or
infinite, such that $\U_1U(t/2)\U_1^*$ is the group of translation to the left by $t$,
and $\U_1V(s)\U_1^*$ is the group of multiplication by $\e^{is\lambda}$. In terms of
the generators, this implies the following. We have
$$
\U_1h(P)\U_1^*=\lambda,
$$
where ``$\lambda$" stands for the multiplication operator by $\lambda$ in
$\ltwo(\R;\C^N,\d\lambda)$, and we have for each $\varphi\in\H(X)$ and
$\phi\in\D_1^0(X)$
\begin{equation}\label{marinera}
\langle\varphi,A\phi\rangle
=\int_\R\d\lambda\,\textstyle\big\langle(\U_1\varphi)(\lambda),
-2i\,\frac{\d(\U_1\phi)}{\d\lambda}(\lambda)\big\rangle_{\C^N},
\end{equation}
where $\frac{\d}{\d\lambda}$ denotes the distributional derivative.

For instance, in the case of the one-dimensional Friedrichs model ($h(x)=x$), one has
$N=1$, and $\U_1$ reduces to the one-dimensional Fourier transform.

{\bf Case 2:} Suppose that $h$ is radial and satisfies the hypotheses of Corollary
\ref{radial_case}. Then there exists a function $h_0\in C^3(\R;\R)$ such that $h(x)=h_0(|x|)$ for each $x\in X$, and we have
$$
\kappa_0:=\kappa(h)=\{\lambda\in\R\mid\exists\rho\in[0,\infty)\textrm{ such that }
h_0(\rho)=\lambda\textrm{ and }h_0'(\rho)=0\}.
$$
In particular $\kappa_0$ is closed as $\kappa(h)$, and it has measure zero due to
Sard's Theorem in $\R$. We also assume that $h_0'\ge0$ on $[0,\infty)$ (so that $h_0^{-1}(\lambda)$ is unique for each $\lambda\in h_0([0,\infty))\setminus\kappa_0$)
and that $h_0^{-1}(\kappa_0)$ has measure zero. These assumptions are satisfied by many physical Hamiltonians such as the Schr\"odinger operator ($h_0(\rho)=\rho^2$) or the square-root Klein-Gordon operator ($h_0(\rho)=\sqrt{1+\rho^2}$).

Taking advantage of the spherical coordinates, one can derive a spectral transformation $\U_0$ for $h(P)\equiv h_0(|P|)$.

\begin{Lemma}\label{transformation}
Let $h_0$ be as above. Then the mapping $\U_0:\H(X)\to\int_{h_0([0,\infty))}^\oplus\d\lambda\,\ltwo(\mathbb S^{d-1})$
defined by
\begin{equation}\label{U_formula}
(\U_0\varphi)(\lambda,\omega)
:=\bigg(\frac{(h_0^{-1}(\lambda))^{d-1}}{h_0'(h_0^{-1}(\lambda))}\bigg)^{1/2}
(\F\varphi)\big(h_0^{-1}(\lambda)\omega\big)
\end{equation}
for each $\varphi\in\H(X)$, $\lambda\in h_0([0,\infty))\setminus\kappa_0$, and
$\omega\in\mathbb S^{d-1}$, is unitary and satisfies
\begin{equation}\label{diagonal}
\U_0h_0(|P|)\U_0^*=\int_{h_0([0,\infty))}^\oplus\d\lambda\,\lambda.
\end{equation}
Moreover, one has for each $\varphi\in\H(X)$ and $\phi\in\D_1^0(X)$
\begin{equation}\label{derivative}
\langle\varphi,A_0\phi\rangle=\int_{h_0([0,\infty))}\d\lambda\,\textstyle
\big\langle(\U_0\varphi)(\lambda,\cdot),
-2i\,\frac{\d(\U_0\phi)}{\d\lambda}(\lambda,\cdot)\big\rangle_{\ltwo(\mathbb S^{d-1})},
\end{equation}
where $\frac\d{\d\lambda}$ denotes the distributional derivative.
\end{Lemma}

Note that Formula \eqref{diagonal} (or the fact that $h_0^{-1}(\kappa_0)$ has measure zero) implies that $h(P)=h_0(|P|)$ has purely absolutely continuous spectrum.
In the case $h_0(\rho)=\rho^2$, $\U_0$ reduces to the usual spectral
transformation for the Schr\"odinger operator \cite[Sec. 2]{Jen81}:
$$
(\U_0\varphi)(\lambda,\omega)
=2^{-1/2}\lambda^{(d-2)/4}(\F\varphi)(\lambda^{1/2}\omega).
$$

\begin{proof}
A direct calculation using the spherical coordinates and the fact that $\kappa_0$
and $h_0^{-1}(k_0)$ have measure zero shows that $\|\U_0\varphi\|^2=\|\varphi\|^2$ for each
$\varphi\in\H(X)$. Thus $\U_0$ is an isometry. Furthermore, for each
$\psi\in\int_{h_0([0,\infty))}^\oplus\d\lambda\,\ltwo(\mathbb S^{d-1})$
and $\xi\in X\setminus\{0\}$, one can check that
\begin{equation}\label{inverse}
\U_0^*\psi=\F^{-1}\widetilde\psi\qquad{\rm where}\qquad
\widetilde\psi(\xi)
:=\bigg(\frac{h_0'(|\xi|)}{|\xi|^{d-1}}\bigg)^{1/2}
\textstyle\psi\big(h_0(|\xi|),\frac\xi{|\xi|}\big).
\end{equation}
Thus $\U_0\U_0^*=1$, and $\U_0$ is unitary. Formulas \eqref{diagonal} and \eqref{derivative}
follow by using \eqref{U_formula}, \eqref{inverse}, and the definition \eqref{A_0}
of $A_0$.
\end{proof}

Formulas \eqref{marinera} and \eqref{derivative} provide (at least when $h$ radial or
a polynomial of degree $1$) a rigorous meaning
to the r.h.s. of Formula \eqref{energy}. They imply that $A$ acts in the spectral representation of $h(P)$ as $-2i\frac\d{\d\lambda}$, where $\lambda$ is the
spectral variable. What about the l.h.s. of Formula \eqref{energy}? For $r$ fixed,
it can be interpreted as the difference of
times spent by the evolving state $\e^{-ith(P)}\varphi$ in the past ($t\le0$)
and in the future ($t\ge0$) within the region defined by the localization operator
$f(Q/r)$. Thus, Formula \eqref{energy} shows (at least when $h$ radial or
a polynomial of degree $1$) that this difference of times tends as $r\to\infty$ to the expectation value
in $\varphi$ of the operator $-2i\frac\d{\d\lambda}$ in the spectral representation
of $h(P)$.

\section{Time delay}\label{time_delay}
\setcounter{equation}{0}

In this section we prove the existence of time delay for scattering systems with free
Hamiltonian $H_0=h(P)$ and full Hamiltonian $H$. The function $h:X\to\R$ satisfies
Assumption \ref{h}, and the full Hamiltonian $H$ can be any selfadjoint operator in
$\H(X)$ satisfying Assumption \ref{wave} below. Given two Hilbert spaces $\H_1$ and $\H_2$,
we write $\B(\H_1,\H_2)$ for the set of bounded operators from $\H_1$ to $\H_2$, and
put $\B(\H_1):=\B(\H_1,\H_1)$. The definition of complete wave operators is given in
\cite[Sec. XI.3]{RSIII}.

\begin{Assumption}\label{wave}
The wave operators $W_\pm$ exist and are complete, and any operator
$T\in\B\big(\H_{-\rho}(X),\H(X)\big)$, with $\rho>\12$, is locally $H$-smooth on
$\R\setminus\{\kappa(h)\cup\sigma_{\rm pp}(H)\}$.
\end{Assumption}

Under Assumption \ref{h} it is known that each operator $T\in\B\big(\H_{-\rho}(X),\H(X)\big)$,
with $\rho>\12$, is locally $h(P)$-smooth on $\R\setminus\kappa(h)$ (see
\cite[Prop. 7.6.6]{ABG} and \cite[Thm. 3.4.3.(a)]{ABG}).
Therefore, if $r>0$ and $\varphi\in\D^0_0(X)$, then $T^0_r(\varphi)$
is finite for each function $f$ satisfying Assumption \ref{assumption_f}.(i) with
$\rho>1$. The number $T_r(\varphi)$ is finite under similar conditions. Indeed,
define for each $t\ge0$
$$
\D_t(X):=\big\{\varphi\in\H_t(X)\mid\eta(h(P))\varphi=\varphi\textrm{ for some }\eta\in
C^\infty_{\rm c}\big(\R\setminus\{\kappa(h)\cup\sigma_{\rm pp}(H)\}\big)\big\}.
$$
Then $T_r(\varphi)$, with $\varphi\in\D_0(X)$, is finite for each
function $f$ satisfying Assumption \ref{assumption_f}.(i) with $\rho>1$ due to
Assumption \ref{wave}. Obviously, the set $\D_t(X)$ satisfies properties similar
to those of $\D^0_t(X)$: $\D_t(X)\subset\H_{\rm ac}(H_0)$, $\D_t(X)$ is dense in
$\H(X)$ if $h^{-1}(\kappa(h)\cup\sigma_{\rm pp}(H))$ has measure zero,
and $\D_{t_1}(X)\subset\D_{t_2}(X)$ if $t_1\ge t_2$.

For each $r>0$, we define the number
\begin{equation}\label{tau_free}
\tau^{\rm free}_r(\varphi)
:=\12\int_0^\infty\d t\<\varphi,S^*\big[\e^{itH_0}f(Q/r)\e^{-itH_0}
-\e^{-itH_0}f(Q/r)\e^{itH_0},S\big]\varphi\>,
\end{equation}
which is finite for all $\varphi\in\D^0_0(X)$. We refer the reader to
\cite[Eq. (93) \& (96)]{AJ07}, \cite[Eq. (4.1)]{GT07}, and \cite[Sec. 2.1]{Tie06} for
similar definitions when $H_0$ is the free Schr\"odinger operator. The usual
definition can be found in \cite[Eq. (3)]{AC87}, \cite[Eq. (6.2)]{Jen81}, and
\cite[Eq. (5)]{Mar76}. The symbol $\R_\pm$ stands for $\R_\pm:=\{x\in\R\mid\pm x\ge0\}$.

\begin{Lemma}\label{lemma_free}
Let $f\ge0$ satisfy Assumption \ref{assumption_f} with $\rho>1$. Suppose that
Assumption \ref{wave} holds. Let $\varphi\in\D_0(X)$ be such that
\begin{equation}\label{H-}
\left\|(W_--1)\e^{-itH_0}\varphi\right\|\in\lone(\R_-,\d t)
\end{equation}
and
\begin{equation}\label{H+}
\left\|(W_+-1)\e^{-itH_0}S\varphi\right\|\in\lone(\R_+,\d t).
\end{equation}
Then
$$
\lim_{r\to\infty}\[\tau_r(\varphi)-\tau^{\rm free}_r(\varphi)\]=0.
$$
\end{Lemma}

\begin{proof}
One has for $\varphi\in\D_0(X)$
\begin{align}
\tau_r(\varphi)-\tau^{\rm free}_r(\varphi)
=&\int_0^\infty\d t\,\Big[\big\|f(Q/r)^{1/2}\e^{-itH}W_-\varphi\big\|^2
-\big\|f(Q/r)^{1/2}\e^{-itH_0}S\varphi\big\|^2\Big]\nonumber\\
&+\int_{-\infty}^0\d t\,\Big[\big\|f(Q/r)^{1/2}\e^{-itH}W_-\varphi\big\|^2
-\big\|f(Q/r)^{1/2}\e^{-itH_0}\varphi\big\|^2\Big].\label{difference}
\end{align}
Using the inequality
$$
\left|\|\varphi\|^2-\|\phi\|^2\right|
\le\|\varphi-\phi\|\cdot\(\|\varphi\|+\|\phi\|\),\quad\varphi,\phi\in\H(X),
$$
the completeness of $W_\pm$, and the fact that $\varphi\in\H_{\rm ac}(H_0)$,
we obtain the estimates
\begin{align}
\Big|\big\|f(Q/r)^{1/2}\e^{-itH}W_-\varphi\big\|^2
-\big\|f(Q/r)^{1/2}\e^{-itH_0}\varphi\big\|^2\Big|
&\le{\rm Const.}\,g_-(t)\,\|\varphi\|\label{borne1}\\
\Big|\big\|f(Q/r)^{1/2}\e^{-itH}W_-\varphi\big\|^2
-\big\|f(Q/r)^{1/2}\e^{-itH_0}S\varphi\big\|^2\Big|
&\le{\rm Const.}\,g_+(t)\,\|\varphi\|\label{borne2},
\end{align}
where
$$
g_-(t):=\left\|(W_--1)\e^{-itH_0}\varphi\right\|\quad{\rm and}\quad
g_+(t):=\left\|(W_+-1)\e^{-itH_0}S\varphi\right\|.
$$
Since $\slim_{r\to\infty}f(Q/r)^{1/2}=1$, the scalars on the l.h.s. of
\eqref{borne1}-\eqref{borne2} converge to $0$ as $r\to\infty$. Furthermore we know from Hypotheses \eqref{H-}-\eqref{H+} that $g_\pm\in\lone(\R_\pm,\d t)$.
Therefore the claim follows from \eqref{difference} and Lebesgue's dominated convergence theorem.
\end{proof}

Next Theorem shows the existence of symmetrized time delay. It is a direct consequence
of Lemma \ref{lemma_free}, Definition \eqref{tau_free}, and Theorem \ref{for_Schwartz}.

\begin{Theorem}\label{sym_case}
Let $f\ge0$ be an even function in $\S(X)$ such that $f=1$ on a bounded neighbourhood
of $0$. Let $h$ satisfy Assumption \ref{h} with $m\ge3$. Suppose that Assumption
\ref{wave} holds. Let $\varphi\in\D_2(X)$ satisfy $S\varphi\in\D_2(X)$ and
\eqref{H-}-\eqref{H+}. Then one has
\begin{equation}\label{Eisenbud_sym}
\lim_{r\to\infty}\tau_r(\varphi)=\12\<\varphi,S^*[A_f,S]\varphi\>,
\end{equation}
with $A_f$ defined by \eqref{A_f}.
\end{Theorem}

\begin{Remark}\label{rem_sym}
The result of Theorem \ref{sym_case} is of particular interest when the localization
function $f$ is radial. In such a case $A_f=A$ due to Lemma \ref{function_R}.(c),
and \eqref{Eisenbud_sym} reduces to
\begin{equation}\label{Eisenbud_radial}
\lim_{r\to\infty}\tau_r(\varphi)
=\12\<\varphi,S^*[A,S]\varphi\>.
\end{equation}
Since $A$ is formally equal to $-2i\frac\d{\d H_0}$, this equation expresses the identity
of symmetrized time delay (defined in terms of sojourn times) and Eisenbud-Wigner time
delay for dispersive Hamiltonians $H_0=h(P)$. To show this more rigorously, let us suppose that $H_0$ is purely absolutely continuous. In such a case there exist Hilbert spaces
$\{\H_\lambda\}_{\lambda\in\sigma(H_0)}$ and a unitary operator $\U:\H(X)\to\int_{\sigma(H_0)}^\oplus\d\lambda\,\H_\lambda$
such that $\U H_0\U^*=\int_{\sigma(H_0)}^\oplus\d\lambda\,\lambda$ and
$\U S\U^*=\int_{\sigma(H_0)}^\oplus\d\lambda\,S(\lambda)$, with $S(\lambda)$ unitary in $\H_\lambda$ (see \eg \cite[Prop. 5.29]{AJS77}). Assume, by analogy to
\eqref{marinera} and \eqref{derivative}, that $A$ satisfies for each
$\varphi\in\H(X)$ and $\phi\in\D_1^0(X)$
\begin{equation}\label{providencia}
\langle\varphi,A\phi\rangle=\int_{\sigma(H_0)}\d\lambda\,\textstyle
\big\langle(\U\varphi)(\lambda),
-2i\,\frac{\d(\U\phi)}{\d\lambda}(\lambda)\big\rangle_{\H_\lambda}.
\end{equation}
Assume also
that the scattering matrix $\sigma(H_0)\ni\lambda\mapsto S(\lambda)\in\H_\lambda$
is strongly continuously differentiable on the support of $\U\varphi$.
Then \eqref{Eisenbud_radial} can be rewritten as
$$
\lim_{r\to\infty}\tau_r(\varphi)
=\int_{\sigma(H_0)}\d\lambda\,\textstyle\big\langle(\U\varphi)(\lambda),-iS(\lambda)^*
\frac{\d S(\lambda)}{\d\lambda}\,(\U\varphi)(\lambda)\big\rangle_{\H_\lambda}.
$$
\end{Remark}

\begin{Remark}\label{two_parts}
One can put into evidence Eisenbud-Wigner contribution to symmetrized
time delay even if the localization function $f$ is not radial. Indeed, by using Formula
\eqref{logarithm}, one gets that
$
A_f=A+\widetilde{A_f},
$
where
$$
\widetilde{A_f}:=Q\cdot\widetilde R_f'(h'(P))
+\widetilde R_f'(h'(P))\cdot Q
$$
and
$$
\textstyle\widetilde R_f(x):=R_f\big(\frac x{|x|}\big)
$$
for each $x\in X\setminus\{0\}$. Thus Formula \eqref{Eisenbud_sym} always implies that
$$
\lim_{r\to\infty}\tau_r(\varphi)
=\12\<\varphi,S^*[A,S]\varphi\>
+\12\big\langle\varphi,S^*[\widetilde{A_f},S]\varphi\big\rangle.
$$
As noted in Remark \ref{rem_sym}, the first term corresponds to the usual Eisenbud-Wigner
time delay. The second term corresponds to the contribution of the non-radial component of the localization function $f$. Due to Equation \eqref{minusone}, one has
$$
\e^{itH_0}\widetilde{A_f}\e^{-itH_0}=\widetilde{A_f}
$$
for each $\varphi\in\D^0_1(X)$ and $t\in\R$. Basically,
this means that $\widetilde{A_f}$ (and thus $S^*[\widetilde{A_f},S]$) is decomposable
in the spectral representation of $H_0$. If $h$ is radial and satisfies the hypotheses
of Lemma \ref{transformation}, one can even determine the restriction
$\widetilde{A_f}(\lambda)$ to the fiber at energy $\lambda$
by using the spectral transformation $\U_0$
($\widetilde{A_f}(\lambda)$ is a symmetric first order differential operator on
$\mathbb S^{d-1}$ with non-constant coefficients).
So, if we sum up, the operator $S^*[\widetilde{A_f},S]$ is
always decomposable in the spectral representation of $H_0$ under some technical
assumptions, but its restriction to the fiber at energy $\lambda$ is an
operator much more complicated than
$-iS(\lambda)^*\frac{\d S(\lambda)}{\d\lambda}$.
Some informations on this matter can be found in \cite[Sec. D]{GT07} in the
particular case of the Schr\"odinger operator ($h(x)=x^2$).
\end{Remark}

Now, we give conditions under which one has
\begin{equation}\label{free_difference}
\lim_{r\to\infty}\[T^0_r(S\varphi)-T^0_r(\varphi)\]=0.
\end{equation}
This implies the equality of time delay and symmetrized time delay in the following
sense:
$$
\lim_{r\to\infty}\[\tau_r^{\rm in}(\varphi)-\tau_r(\varphi)\]=0.
$$
Physically, \eqref{free_difference} means that the freely evolving states
$\e^{-itH_0}\varphi$ and $\e^{-itH_0}S\varphi$ tend to spend the same time within
the region defined by the localization function $f(Q/r)$ as $r\to\infty$.
Formally, the proof of
\eqref{free_difference} goes as follows. Suppose that $F_f(h'(P))$, with $F_f$ defined
in Section \ref{averaged}, commutes with the scattering operator $S$. Then, using the
change of variables $\mu:=t/r$, $\nu:=1/r$, and the parity of $f$, one gets
\begin{align*}
\lim_{r\to\infty}\[T^0_r(S\varphi)-T^0_r(\varphi)\]
&=\lim_{r\to\infty}\int_\R\d t\,
\big\langle\varphi,S^*[\e^{ith(P)}f(Q/r)\e^{-ith(P)},S]\varphi\big\rangle\\
&\qquad-\<\varphi,S^*[F_f(h'(P)),S]\varphi\>\\
&=\lim_{\nu\searrow0}\int_\R\d\mu\,
\big\langle\varphi,S^*\big[\textstyle\frac1\nu
\big\{f(\nu Q+\mu h'(P))-f(\mu h'(P))\big\},S\big]\varphi\big\rangle\\
&=\int_\R\d\mu\,
\big\langle\varphi,S^*[Q\cdot f'(\mu h'(P)),S]\varphi\big\rangle\\
&=0.
\end{align*}
The rigorous proof will be given in Theorem
\ref{equal_sojourn} below. Before this we introduce assumptions on $h$ slightly
stronger than Assumption \ref{h}, and we prove a technical lemma.

\begin{Assumption}\label{h_strong}
The function $h:X\to\R$ is of class $C^m$ for some $m\ge2$, and satisfies the
following conditions:
\begin{enumerate}
\item[(i)] $|h(x)|\to\infty$~~as~~$|x|\to\infty$.
\item[(ii)] $\sum_{|\alpha|\le m}|(\partial^\alpha h)(x)|\le{\rm Const.}\;\!(1+|h(x)|)$.
\item[(iii)] $\sum_{|\alpha|=m}|(\partial^\alpha h)(x)|\le{\rm Const.}$
\end{enumerate}
\end{Assumption}

Assumption \ref{h_strong} appears naturally when one studies the spectral and scattering theory of pairs $\{H_0=h(P),H\}$ using commutator methods (see \eg \cite[Sec. 7.6.3]{ABG}
and \cite[Sec. 2.1]{Sah97_2}). Assumption
\ref{h_strong}.(i) is related to the closedness of $\kappa(h)$, whereas Assumptions
\ref{h_strong}.(ii)-(iii) are related to the polynomial growth of the group
$\{\e^{ix\cdot Q}\}$ in $\dom(H_0)$ and $\dom(|H_0|^{1/2})$. We say that functions $h$ satisfying Assumption \ref{h_strong} are of hypoelliptic type, by reference to
hypoelliptic polynomials of degree $m$ which also satisfy Assumption \ref{h_strong}
(see \cite[Thm. 11.1.3]{Hor05}). A typical example
one should keep in mind is the case where $h$ is an elliptic symbol of degree $s>0$,
\ie $h\in C^\infty(X;\R)$,
$|(\partial^\alpha h)(x)|\le{\textsc c}_\alpha\langle x\rangle^{s-|\alpha|}$ for
each multi-index $\alpha$, and $|h(x)|\ge{\textsc c}\,|x|^s$, for some
${\textsc c}>0$, outside a compact set. 

\begin{Lemma}\label{batman}
Let $h$ satisfy Assumption \ref{h_strong} with $m\ge2$, and take
$\eta\in C^\infty_{\rm c}\big(\R\setminus\kappa(h)\big)$. Then one has for each
$\mu\in\R$, $x\in X$, and $|\nu|<1$
$$
\big\|\textstyle\frac1\nu\big\{\eta(h(P+\nu x))\e^{i\frac\mu\nu[h(P+\nu x)-h(P)]}
-\eta(h(P))\e^{i\mu x\cdot h'(P)}\big\}\big\|
\le{\rm Const.}\,(1+|\mu|)\<x\>^{m+2}.
$$
\end{Lemma}

\begin{proof}
Due to the spectral theorem and the mean value theorem, one has
\begin{equation}
\big\|\textstyle\frac1\nu\big\{\eta(h(P+\nu x))\e^{i\frac\mu\nu[h(P+\nu x)-h(P)]}
-\eta(h(P))\e^{i\mu x\cdot h'(P)}\big\}\big\|
\displaystyle\le\sup_{y\in X,~\xi\in[0,1]}\big|g_y'(\xi\nu)\big|,\label{bibou1}
\end{equation}
where
\begin{align*}
g_y(\nu)&:=\eta(h(y+\nu x))\e^{i\frac\mu\nu[h(y+\nu x)-h(y)]}\\
&=\eta(h(y+\nu x))\exp\big[\textstyle
i\mu\sum_{|\alpha|=1}x^\alpha\int_0^1\d t\,(\partial^\alpha h)(y+t\nu x)\big].
\end{align*}
Direct calculations using Assumption \ref{h_strong}.(ii) show that
\begin{equation}\label{bibou2}
\sup_{\xi\in[0,1]}\big|g_y'(\xi\nu)\big|
\le{\rm Const.}\,|x|+{\rm Const.}\,x^2|\mu|\sup_{\xi,t\in[0,1]}
\big|\eta(h(y+\xi\nu x))\big|\big(1+|h(y+t\xi\nu x)|\big).
\end{equation}
Then one can use Taylor's Formula \cite[Eq. (1.1.8)]{ABG}
\begin{align*}
&h(y+t\xi\nu x)\\
&=\sum_{|\alpha|<m}\frac{[(t-1)\xi\nu]^{|\alpha|}x^\alpha}{\alpha!}
\,(\partial^\alpha h)(y+\xi\nu x)\\
&\qquad+m[(t-1)\xi\nu]^m\sum_{|\alpha|=m}\frac{x^\alpha}{\alpha!}\int_0^1\d\tau\,
(\partial^\alpha h)\big(y+\xi\nu x+\tau(t-1)\xi\nu x\big)(1-\tau)^{m-1}
\end{align*}
to get a bound for $|h(y+t\xi\nu x)|$ in terms of $|h(y+\xi\nu x)|$. Indeed, using
the formula above and Assumptions \ref{h_strong}.(ii)-(iii), one obtains that
$$
|h(y+t\xi\nu x)|
\le{\rm Const.}\,\langle\nu\rangle^{m-1}\langle x\rangle^{m-1}\big(1+|h(y+\xi\nu x)|\big)
+{\rm Const.}\,|\nu|^m|x|^m.
$$
This, together with the bounds
\eqref{bibou1}-\eqref{bibou2} and Assumption \ref{h_strong}.(ii), implies the claim.
\end{proof}

\begin{Theorem}\label{equal_sojourn}
Let $f\in\S(X)$ be even, let $h$ satisfy Assumption \ref{h_strong} with
$m\ge3$, and suppose that Assumption \ref{wave} holds. If $\varphi\in\D_2^0(X)$ satisfies $S\varphi\in\D_2^0(X)$ and
\begin{equation}\label{grobobo}
[F_f(h'(P)),S]\varphi=0,
\end{equation}
then one has
\begin{equation}\label{clementine}
\lim_{r\to\infty}\[T^0_r(S\varphi)-T^0_r(\varphi)\]=0.
\end{equation}
In particular, time delay and symmetrized time delay satisfy
\begin{equation}\label{roblochon}
\lim_{r\to\infty}\[\tau_r^{\rm in}(\varphi)-\tau_r(\varphi)\]=0.
\end{equation}
\end{Theorem}

The l.h.s. in \eqref{grobobo} is well-defined due to Equation \eqref{homo_f}. Indeed,
one has
$$
[F_f(h'(P)),S]\varphi
=\big[|h'(P)|^{-1}\eta(h(P))F_f\big(\textstyle\frac{h'(P)}{|h'(P)|}\big),S\big]\varphi
$$
for some $\eta\in C^\infty_{\rm c}\big(\R\setminus\kappa(h)\big)$, and thus
$[F_f(h'(P)),S]\varphi\in\H(X)$ by \eqref{bo-bound} and the compacity of
$F_f(\mathbb S^{d-1})$.

\begin{proof}
Let $\varphi\in\D_2^0(X)$, take a real
$\eta\in C^\infty_{\rm c}\big(\R\setminus\kappa(h)\big)$ such that $\eta(h(P))\varphi=\varphi$, and set $\eta_t(P):=\e^{ith(P)}\eta(h(P))$. Using \eqref{grobobo} and the change of variables $\mu:=t/r$, $\nu:=1/r$, one gets
\begin{align}
&T^0_{1/\nu}(S\varphi)-T^0_{1/\nu}(\varphi)\nonumber\\
&=\int_\R\d\mu\,\big\langle\varphi,S^*\big[\textstyle\frac1\nu
\big\{\eta_{\frac\mu\nu}(P)f(\nu Q)\eta_{-\frac\mu\nu}(P)
-f(\mu h'(P))\big\},S\big]\varphi\big\rangle\nonumber\\
&=\int_\R\d\mu\int_X\underline\d x\,(\F f)(x)
\big\langle\varphi,S^*\big[\textstyle\frac1\nu\big\{\e^{i\nu x\cdot Q}
\eta_{\frac\mu\nu}(P+\nu x)\eta_{-\frac\mu\nu}(P)
-\e^{i\mu x\cdot h'(P)}\big\},S\big]\varphi\big\rangle\nonumber\\
&=\int_\R\d\mu\int_X\underline\d x\,(\F f)(x)
\big\langle\varphi,S^*\big[\textstyle\frac1\nu(\e^{ivx\cdot Q}-1)
\eta(h(P+\nu x))\e^{i\frac\mu\nu[h(P+\nu x)-h(P)]},S\big]\varphi\big\rangle
\label{first_term}\\
&\quad+\int_\R\d\mu\int_X\underline\d x\,(\F f)(x)
\big\langle\varphi,S^*\big[\textstyle\frac1\nu
\big\{\eta(h(P+\nu x))\e^{i\frac\mu\nu[h(P+\nu x)-h(P)]}\nonumber\\
&\hspace{220pt}
-\eta(h(P))\e^{i\mu x\cdot h'(P)}\big\},S\big]\varphi\big\rangle.\nonumber
\end{align}
To prove the statement, it is sufficient to show that the limit as $\nu\searrow0$ of
each of these two terms is equal to zero. This is done in points (i) and (ii) below.

(i) One can adapt the method Theorem \ref{for_Schwartz} (point (iii) of the
proof) in order to apply Lebesgue's dominated convergence theorem to
\eqref{first_term}. So one gets
\begin{align*}
&\lim_{\nu\searrow0}\int_\R\d\mu\int_X\underline\d x\,(\F f)(x)
\big\langle\varphi,S^*\big[\textstyle\frac1\nu(\e^{ivx\cdot Q}-1)
\eta(h(P+\nu x))\e^{i\frac\mu\nu[h(P+\nu x)-h(P)]},S\big]\varphi\big\rangle\\
&=i\int_\R\d\mu\int_X\underline\d x\,(\F f)(x)
\big\{\big\langle(x\cdot Q)S\varphi,\e^{i\mu x\cdot h'(P)}S\varphi\big\rangle
-\big\langle(x\cdot Q)\varphi,\e^{i\mu x\cdot h'(P)}\varphi\big\rangle\big\},
\end{align*}
and the change of variables $\mu':=-\mu$, $x':=-x$, together with the parity of $f$,
implies that this expression is equal to zero.

(ii) We have to show that the limit
\begin{align}
\ell&:=\lim_{\nu\searrow0}\int_\R\d\mu\int_X\underline\d x\,(\F f)(x)
\big\langle\varphi,S^*\big[\textstyle\frac1\nu
\big\{\eta(h(P+\nu x))\e^{i\frac\mu\nu[h(P+\nu x)-h(P)]}\label{second_term}\\
&\hspace{220pt}
-\eta(h(P))\e^{i\mu x\cdot h'(P)}\big\},S\big]\varphi\big\rangle\nonumber
\end{align}
is equal to zero. For the moment, let us assume that we can interchange the
limit and the integrals
in \eqref{second_term}, by invoking Lebesgue's dominated convergence theorem. Since
\begin{align*}
&\big\{\textstyle\frac\d{\d\nu}\,\eta(h(P+\nu x))
\e^{i\frac\mu\nu[h(P+\nu x)-h(P)]}\big\}_{\nu=0}\\
&=x\cdot h'(P)\eta'(P)\e^{i\mu x\cdot h'(P)}
+{\textstyle\frac{i\mu}2}\,\eta(h(P))\sum_{|\alpha|=2}x^\alpha
(\partial^\alpha h)(P)\e^{i\mu x\cdot h'(P)},
\end{align*}
one gets in such a case
\begin{align*}
\ell&=\int_\R\d\mu\int_X\underline\d x\,(\F f)(x)\big\langle\varphi,
S^*\big[x\cdot h'(P)\eta'(P)\e^{i\mu x\cdot h'(P)},S\big]\varphi\big\rangle\\
&\quad+\textstyle\frac i2\sum_{|\alpha|=2}\int_\R\d\mu\,\mu
\int_X\underline\d x\,x^\alpha(\F f)(x)\big\langle\varphi,
S^*\big[(\partial^\alpha h)(P)\e^{i\mu x\cdot h'(P)},S\big]\varphi\big\rangle.
\end{align*}
Then the change of variables $\mu':=-\mu$, $x':=-x$, together with the parity of $f$,
implies that this expression is equal to zero.

It remains to show that one can apply Lebesgue's dominated convergence theorem to
\eqref{second_term}. Since $\varphi$ and $S\varphi$ belong to the same set $\D_2^0(X)$
it is enough to treat the limit $\lim_{\nu\searrow0}\int_\R\d\mu\,L(\nu,\mu)$, where
\begin{align*}
L(\nu,\mu):=\int_X\underline\d x\,(\F f)(x)
\big\langle\varphi,\textstyle\frac1\nu\big\{\eta(h(P+\nu x))&
\e^{i\frac\mu\nu[h(P+\nu x)-h(P)]}\\
&-\eta(h(P))\e^{i\mu x\cdot h'(P)}\big\}
\varphi\big\rangle.
\end{align*}
Using Lemma \ref{batman} and the fact that $\F f\in\S(X)$, one gets that
$|L(\nu,\mu)|\le{\rm Const.}\,(1+|\mu|)$ for all $|\nu|<1$. Therefore $L(\nu,\mu)$ is
bounded uniformly for $|\nu|<1$ by a function in $\lone([-1,1],\d\mu)$.

For the case
$|\mu|>1$ we recall that there exists $\textsc c>0$ such that $|h'(x)|>\textsc c$ for
all $x\in h^{-1}(\supp\eta)$, due to Assumption \ref{h_strong}.(i). So $L(\nu,\mu)$ can be
rewritten as
\begin{align*}
L(\nu,\mu)=\sum_{j\le d}\int_X\underline\d x\,(\F f)(x)
\big\langle\varphi,\textstyle\frac1\nu
\big\{\frac{\eta(h(P+\nu x))(\partial_jh)(P+\nu x)}{i\mu|h'(P+\nu x)|^2}\,
\big(\partial_j\e^{i\frac\mu\nu[h(P+\nu x)-h(P)]}\big)\\
-\textstyle\frac{\eta(h(P))(\partial_jh)(P)}{i\mu|h'(P)|^2}\,
\big(\partial_j\e^{i\mu x\cdot h'(P)}\big)\big\}\varphi\big\rangle,
\end{align*}
and one can perform an integration by parts (with vanishing boundary contributions)
with respect to $x_j$. We do not give the details here since the calculations are
very similar to those of Theorem \ref{for_Schwartz} (point (iii) of the proof). We
only give the result obtained after three successive integrations by parts:
\begin{align}
&L(\nu,\mu)=\O(|\mu|^{-2})-i\mu^{-3}\sum_{j,k\le d}\int_X\underline\d x\,
[\partial_k\partial_j^2(\F f)(x)]~\times\label{choclo}\\
&\times~\big\langle\varphi,\textstyle\frac1\nu
\big\{\frac{\eta(h(P+\nu x))(\partial_kh)(P+\nu x)}{|h'(P+\nu x)|^4}\,
\e^{i\frac\mu\nu[h(P+\nu x)-h(P)]}
-\frac{\eta(h(P))(\partial_kh)(P)}{|h'(P)|^4}\,
\e^{i\mu x\cdot h'(P)}\big\}\varphi\big\rangle,\nonumber
\end{align}
where $\O(|\mu|^{-2})$ are terms (containing derivatives $\partial^\alpha h$ with $|\alpha|\le3$) bounded in norm by ${\rm Const.}\,|\mu|^{-2}$. Now, one shows as in
Lemma \ref{batman} that
\begin{align*}
&\big\|\textstyle\frac1\nu
\big\{\frac{\eta(h(P+\nu x))(\partial_kh)(P+\nu x)}{|h'(P+\nu x)|^4}\,
\e^{i\frac\mu\nu[h(P+\nu x)-h(P)]}
-\frac{\eta(h(P))(\partial_kh)(P)}{|h'(P)|^4}\,
\e^{i\mu x\cdot h'(P)}\big\}\big\|\\
&\le{\rm Const.}\,(1+|\mu|)\<x\>^{m+2}
\end{align*}
for each $\mu\in\R$, $x\in X$, and $|\nu|<1$. It follows by \eqref{choclo} that $|L(\nu,\mu)|\le{\rm Const.}\,|\mu|^{-2}$ for each
$|\nu|<1$. This bound, together
with our previous estimate for $|\mu|\le1$, showns that $L(\nu,\mu)$ is bounded
uniformly for $|\nu|<1$ by a function in $\lone(\R,\d\mu)$. So one can interchange
the limit $\nu\searrow0$ and the integration over $\mu$ in \eqref{second_term}.

The interchange of the limit $\nu\searrow0$ and the integration over $x$ in \eqref{second_term} is justified by the bound
\begin{align*}
&\big|(\F f)(x)\big\langle\varphi,\textstyle\frac1\nu\big\{\eta(h(P+\nu x))
\e^{i\frac\mu\nu[h(P+\nu x)-h(P)]}-\eta(h(P))\e^{i\mu x\cdot h'(P)}\big\}
\varphi\big\rangle\big|\\
&\le{\rm Const.}\,(1+|\mu|)\big|(\F f)(x)\big|\langle x\rangle^{m+2},
\end{align*}
which follows from Lemma \ref{batman}.
\end{proof}

In physical terms, the commutation condition \eqref{grobobo}
expresses roughly the conservation of the observable $F_f(h'(P))$ by the scattering
process. Since $h'(P)$ is the free velocity operator for the scattering
process, $F_f(h'(P))$ is a quantum analogue of the classical sojourn
time $F_f(p)$, with momentum $p\in\R$, described at the end of Section
\ref{averaged}. Therefore it is not completely surprising that the sojourn times $T^0_r(S\varphi)$
and $T^0_r(\varphi)$ are equal (in the sense of \eqref{clementine}) if
\eqref{grobobo} is satisfied.

\begin{Remark}\label{rem_com}
There are many situations where the commutation assumption \eqref{grobobo}
is satisfied. Here we present two of them. The first one occurs when $h$ is a polynomial of degree $1$, \ie $h(x)=v_0+v\cdot x$ for some
$v_0\in\R$, $v\in X\setminus\{0\}$. In such a case the operator
$F_f(h'(P))$ reduces to the scalar $F_f(v)$, and thus \eqref{grobobo} is clearly
satisfied. The second one occurs when both $f$ and $h$ are radial, namely when
$f(x)=f_0(|x|)$ and $h(x)=h_0(|x|)$ with, say, $h_0$ as in Lemma \ref{transformation}.
In such a case $F_f(h'(P))$ is diagonalizable in the spectral
representation of $H_0\equiv h(P)$, namely
\begin{equation}\label{F_diag}
\U_0F_f(h'(P))\U_0^*=\int_{h_0([0,\infty))}^\oplus\d\lambda\,
F_f\big(h_0'(h_0^{-1}(\lambda))\big),
\end{equation}
where $\U_0$ is the spectral transformation \eqref{U_formula} for $h(P)$.
We also know, under Assumption \ref{wave}, that $S$ is decomposable
in the spectral representation of $H_0$. Thus
\eqref{grobobo} is satisfied, since diagonalizable operators commute with
decomposable operators.
\end{Remark}

We are now in a position to state our main theorem on the existence of time delay.
It is a direct consequence of Theorems \ref{sym_case} and \ref{equal_sojourn}.

\begin{Theorem}\label{big_one}
Let $f\ge0$ be an even function in $\S(X)$ such that $f=1$ on a bounded neighbourhood of
$0$. Let $h$ satisfy Assumption \ref{h_strong} with $m\ge3$. Suppose that Assumption \ref{wave} holds. Let $\varphi\in\D_2(X)$ satisfy $S\varphi\in\D_2(X)$, \eqref{grobobo},
and \eqref{H-}-\eqref{H+}. Then one has
\begin{equation}\label{main}
\lim_{r\to\infty}\tau_r^{\rm in}(\varphi)
=\lim_{r\to\infty}\tau_r(\varphi)
=\12\<\varphi,S^*[A_f,S]\varphi\>,
\end{equation}
with $A_f$ defined by \eqref{A_f}.
\end{Theorem}

The comments of Remarks \ref{rem_sym} and \ref{two_parts} concerning the
symmetrized time delay $\tau_r(\varphi)$ remain valid in the case of the
time delay $\tau_r^{\rm in}(\varphi)$.
The r.h.s. of \eqref{main} can always be written as the sum of the
Eisenbud-Wigner time delay and the time delay associated to the
non-radial component of the localization function $f$. In particular, if $f$ is radial,
one has
\begin{equation}\label{lennie}
\lim_{r\to\infty}\tau_r^{\rm in}(\varphi)
=\int_{\sigma(H_0)}\d\lambda\,\textstyle\big\langle(\U\varphi)(\lambda),-iS(\lambda)^*
\frac{\d S(\lambda)}{\d\lambda}\,(\U\varphi)(\lambda)\big\rangle_{\H_\lambda}
\end{equation}
under the assumptions of Remark \ref{rem_sym}.

Formula \eqref{lennie} is the main result of this paper: it expresses the
identity of time delay (defined in terms of sojourn times) and Eisenbud-Wigner
time delay for dispersive Hamiltonians $H_0=h(P)$. However, \eqref{lennie} holds
only if the conditions \eqref{providencia} and \eqref{grobobo} are satisfied.
As we have seen in cases $1$ and $2$ of Section \ref{integral} and Remark
\ref{rem_com}, this occurs for instance when $h$ is a polynomial of degree $1$
or radial. These two classes of functions
provide a bulk of examples much bigger than what can be found in the literature,
since only the Schr\"odinger Hamiltonian ($h(x)=x^2$) have been explicitly
treated before.

We collect the preceding remarks in a corollary to Theorem \ref{big_one}.

\begin{Corollary}\label{two_cases}
Let $f\ge0$ be an even function in $\S(X)$ such that $f=1$ on a bounded neighbourhood
of $0$. Let $h$ satisfy Assumption \ref{h_strong} with $m\ge3$. Suppose that Assumption
\ref{wave} holds. Let $\varphi\in\D_2(X)$ satisfy $S\varphi\in\D_2(X)$ and
\eqref{H-}-\eqref{H+}. Then
\begin{enumerate}
\item[(a)] Suppose that $h(x)=v_0+v\cdot x$ for some $v_0\in\R$,
$v\in X\setminus\{0\}$. Then one has
$$
\lim_{r\to\infty}\tau_r^{\rm in}(\varphi)
=\int_\R\d\lambda\,\textstyle\big\langle(\U_1\varphi)(\lambda),
-iS(\lambda)^*\,\frac{\d S(\lambda)}{\d\lambda}(\U_1\varphi)(\lambda)\big\rangle_{\C^N}
$$
if the scattering matrix $\R\ni\lambda\mapsto S(\lambda)\in\B(\C^N)$
is strongly continuously differentiable on the support of $\U_1\varphi$.
\item[(b)] Let $f$ be radial, and suppose that $h$ is radial and satisfies the
hypotheses of Lemma \ref{transformation}. Then one has
$$
\lim_{r\to\infty}\tau_r^{\rm in}(\varphi)
=\int_{h_0([0,\infty))}\d\lambda\,\textstyle
\big\langle(\U_0\varphi)(\lambda,\cdot),
-iS(\lambda)^*\,\frac{\d S(\lambda)}{\d\lambda}
(\U_0\varphi)(\lambda,\cdot)\big\rangle_{\ltwo(\mathbb S^{d-1})}
$$
if the scattering matrix $h_0([0,\infty))\ni\lambda\mapsto
S(\lambda)\in\B\big(\ltwo(\mathbb S^{d-1})\big)$ is strongly continuously
differentiable on the support of $\U_0\varphi$.
\end{enumerate}
\end{Corollary}

\section{Friedrichs model}\label{Friedrichs}
\setcounter{equation}{0}

As an illustration of our results, we treat in this section the case of a one-dimensionnal Friedrichs Hamiltonian $H_0$ perturbed by a
finite rank operator $V$. For historical reasons \cite{Fri38} we define the Friedrichs Hamiltonian as the position operator $H_0:=Q$ in the Hilbert space $\H(\R):=\ltwo(\R)$.
The operator $H_0$ satisfies $\F H_0\F^{-1}=-P$. So, we can apply after a Fourier transformation the results of the Section \ref{time_delay} with $h(x)=-x$ and $\kappa(h)=\varnothing$. Since $h$ is a polynomial of degree $1$, we only
have to check the hypotheses of Corollary \ref{two_cases}.(a) in order to prove the
existence of the limits $\lim_{r\to\infty}\tau_r^{\rm in}(\varphi)$ and $\lim_{r\to\infty}\tau_r(\varphi)$, and their identity with Eisenbud-Wigner time delay.
However, the model is very explicit, so we will add some more remarks to this result.

\subsection{Preliminaries}\label{Preliminaries}

For the moment, we do not specify the selfadjoint perturbation $H$ of $H_0=Q$. We only
assume, by analogy to Assumption \ref{wave}, that

\begin{Assumption}\label{wave_F}
The wave operators $W_\pm$ exist and are complete, and any operator
$T\in\B\big(\H^{-s}(\R),\H(\R)\big)$, with $s>\12$, is locally $H$-smooth on
$\R\setminus\sigma_{\rm pp}(H)$.
\end{Assumption}

Since $H_0=Q$ the propagation of the states $\varphi\in\H(\R)$ takes place in the
space of momenta. Therefore the quantities $T^0_r(\varphi)$, $T_r(\varphi)$,
$\tau_r^{\rm in}(\varphi)$, and $\tau_r(\varphi)$ are defined with respect to a
localization operator $f(P/r)$:
\begin{align*}
T^0_r(\varphi)&:=\int_\R\d t\<\e^{-itH_0}\varphi,f(P/r)\e^{-itH_0}\varphi\>,\\
T_r(\varphi)&:=\int_\R\d t\<\e^{-itH}W_-\varphi,f(P/r)\e^{-itH}W_-\varphi\>,\\
\tau_r^{\rm in}(\varphi)&:=T_r(\varphi)-T^0_r(\varphi),\\
\tau_r(\varphi)&:=T_r(\varphi)-\12\[T^0_r(\varphi)+T^0_r(S\varphi)\].
\end{align*}
The sets $\D^0_t(X)$ and $\D_t(X)$ of Sections \ref{integral} and \ref{time_delay}
are replaced by
$$
\D^s_0(\R):=\big\{\varphi\in\H^s(\R)\mid\eta(Q)\varphi=\varphi
\textrm{ for some }\eta\in C^\infty_{\rm c}(\R)\big\}
$$
and 
$$
\D^s(\R):=\big\{\varphi\in\H^s(\R)\mid\eta(Q)\varphi=\varphi\textrm{ for some }
\eta\in C^\infty_{\rm c}\big(\R\setminus\sigma_{\rm pp}(H)\big)\big\}
$$
for $s\ge0$. Theorem \ref{for_Schwartz} implies that
\begin{equation}\label{pimpam}
\lim_{r\to\infty}\int_0^\infty\d t\,\big\langle\varphi,\big[\e^{itQ}f(P/r)\e^{-itQ}
-\e^{-itQ}f(P/r)\e^{itQ}\big]\varphi\big\rangle=2\big\langle\varphi,P\varphi\big\rangle
\end{equation}
for each $\varphi\in\D_0^2(\R)$ and each even function $f\in\S(\R)$ such that $f=1$ on a bounded neighbourhood of $0$. Using the formula
\begin{equation}\label{evolution}
\textstyle\e^{itQ}g(P/r)\e^{-itQ}=g\big(\frac{P-t}r\big),\quad g\in\linf(\R),
\end{equation}
one can even show that \eqref{pimpam} remains true for all $\varphi\in\H^s(\R)$, $s>1$,
and all $f$ satisfying the following assumption.

\begin{Assumption}\label{assumption_Frie}
The function $f\in\linf(\R)$ is even, $f=1$ on a bounded neighbourhood of $0$, and there
exists $\rho>1$ such that $|f(x)|\le{\rm Const.}\<x\>^{-\rho}$ for a.e. $x\in\R$.
\end{Assumption}

The typical example of function $f$ one should keep in mind is the following.

\begin{Example}\label{projection}
Let $f=\chi_J$, where $J\subset\R$ is bounded, symmmetric (\ie $J=-J$), and contains an interval
$(-\delta,\delta)$ for some $\delta>0$. Then $f$ satisfies Assumption \ref{assumption_Frie}, and
$f(P/r)$ is the orthogonal projection onto the set of states with momentum localised in $rJ$.
\end{Example}

Formula \eqref{evolution} and the parity of $f$ give for each $r>0$ and $\varphi\in\H(\R)$
$$
T^0_r(\varphi)
=\int_\R\d t\int_\R\d k\,|(\F\varphi)(k)|^2\textstyle f\big(\frac{t-k}r\big).
$$
Then Fubini's theorem (which is applicable due to Assumption \ref{assumption_Frie}) and
the change of variable $x:=\frac{t-k}r$ imply that
\begin{equation}\label{papate}
T^0_r(\varphi)=r\|\varphi\|^2\int_\R\d x\,f(x),
\end{equation}
and thus that
\begin{equation}\label{doduffe}
T^0_r(S\varphi)=T^0_r(\varphi)\qquad{\rm and}\qquad
\tau_r^{\rm in}(\varphi)=\tau_r(\varphi).
\end{equation}
So the equations \eqref{clementine} and \eqref{roblochon} of Theorem
\ref{equal_sojourn} are true here not only as $r\to\infty$, but for each $r>0$.
This can be explained as follows. The ``velocity" operator associated
with the free evolution group $\e^{itQ}$ is not only constant (which guarantees
that Theorem \ref{equal_sojourn} is applicable), but equal to $-1$:
$$
\frac\d{\d t}\,(\e^{itQ}P\e^{-itQ})=-1.
$$
Therefore the propagation speed of a state $\e^{itQ}\varphi$ in the space of momenta is equal to $-1$. In that respect Formulas \eqref{papate}-\eqref{doduffe} are natural. For instance, if
$\|\varphi\|=1$ and $f=\chi_J$ is as in Example \ref{projection}, then
$T^0_r(\varphi)=r|J|$, where $|J|$ is the Lebesgue measure of $J$. So
$T^0_r(\varphi)$ is nothing else but the sojourn time in $rJ$ (in the space of momenta) of the state $\e^{itQ}\varphi$ propagating at speed $-1$.

Next Lemma follows from what precedes and Theorem \ref{big_one}.

\begin{Lemma}\label{abs_delay}
Let $f\ge0$ satisfy Assumption \ref{assumption_Frie}. Suppose that Assumption
\ref{wave_F} holds. For
some $s>1$, let $\varphi\in\D^s(\R)$ satisfy \eqref{H-}-\eqref{H+} and
$S\varphi\in\D^s(\R)$. Then
\begin{equation}\label{com_P}
\lim_{r\to\infty}\tau_r^{\rm in}(\varphi)
=\lim_{r\to\infty}\tau_r(\varphi)
=\<\varphi,S^*[P,S]\varphi\>.
\end{equation}
\end{Lemma}

\begin{Remark}
Formula \eqref{com_P} shows that $\lim_{r\to\infty}\tau^{\rm in}_r(\varphi)$ is null if the commutator
$[P,S]$ vanishes (which happens if and only if the scattering operator $S$ is a constant). We give an
example of Hamiltonian $H$ for which this occurs.

Let $\widetilde{H_0}:=P$ with domain $\dom(\widetilde{H_0}):=\H^1(\R)$, and for $q\in\lone(\R;\R)$
let $\widetilde H:=\widetilde{H_0}+q(Q)$ with domain
$\dom(\widetilde H):=\big\{\varphi\in\H^1(\R)\mid\widetilde H\varphi\in\H(\R)\big\}$. It is known
\cite[Sec. 2.4.3]{Yaf92} that $\widetilde H$ is selfadjoint, that the wave operators
$\widetilde{W_\pm}:=\slim_{s\to\pm\infty}\e^{it\widetilde H}\e^{-it\widetilde{H_0}}$ exist and
are complete, and that $\widetilde S:=\widetilde{W_+}^*\widetilde{W_-}=\e^{-i\int_\R\d x\;\!q(x)}$
is a constant. Therefore $H:=\F\widetilde H\F^{-1}=H_0+q(-P)$ is selfadjoint on
$\dom(H):=\F\dom(\widetilde H)$, the wave operators $W_\pm=\F\widetilde{W_\pm}\F^{-1}$ exist and
are complete, and $S=\widetilde S$.
\end{Remark}

\begin{Remark}\label{Remark_Eisenbud}
Suppose that the assumptions of Lemma \ref{abs_delay} are verified, and for a.e. $x\in\R$ let
$S(x)\in\C$ be the component at energy $x$ of the scattering matrix associated with the scattering operator $S$.
Then Equation \eqref{com_P} can be rewritten as
\begin{equation}\label{Eisenbud}
\lim_{r\to\infty}\tau_r^{\rm in}(\varphi)
=\lim_{r\to\infty}\tau_r(\varphi)
=-i\int_\R\d x\,|\varphi(x)|^2\overline{S(x)}S'(x)
\end{equation}
if the function $x\mapsto S(x)$ is continuously differentiable on the support of $\varphi$
(note that Equation \eqref{Eisenbud} does not follow from \cite{Mar76} or \cite[Chap. 7.2]{AJS77},
since we do not require $f(P/r)$ to be an orthogonal projection or $x\mapsto S(x)$ to be twice
differentiable on the whole real line). Formula \eqref{Eisenbud} holds for the general class of
functions $f\ge0$ satisfying Assumption \ref{assumption_Frie}. However, if $\|\varphi\|=1$ and $f=\chi_J$
is as in Example \ref{projection}, then we know that the scalars $T^0_r(\varphi)$ and $T_r(\varphi)$
can be interpreted as sojourn times. Therefore in such a case Formula \eqref{Eisenbud} expresses exactly
the identity of the usual and symmetrised time delay with the Eisenbud-Wigner time delay
for the Friedrichs model.
\end{Remark}

\begin{Remark}\label{RemarkBirman}
Let $R_0(\cdot)$ and $R(\cdot)$ be the resolvent families of $H_0$ and $H$, and suppose that
$R(i)-R_0(i)$ is trace class. Then, at least formally, we get from the Birman-Krein formula
\cite[Thm. 8.7.2]{Yaf92} that
\begin{equation}\label{too_much}
\overline{S(x)}S'(x)=-2\pi i\xi'(x;H,H_0),
\end{equation}
where $\xi'(x;H,H_0)$ is the derivative of the spectral shift function for the pair $\{H_0,H\}$.
Therefore one has
\begin{equation}\label{too_too}
\lim_{r\to\infty}\tau^{\rm in}_r(\varphi)=-2\pi\int_\R\d x\,|\varphi(x)|^2\xi'(x;H,H_0),
\end{equation}
and the number $-2\pi\xi'(x;H,H_0)$ may be interpreted as the component at energy $x$ of the
time delay operator for the Friedrichs model. However Equations \eqref{too_much}-\eqref{too_too}
turn out to be difficult to prove rigorously under this form. We refer to \cite{JSM72},
\cite[Sec. III.b]{Mar81}, and \cite[Sec. 3]{Rob94} for general theories on this issue, and to \cite{Bus71,Dre78,Oga78,Yaf80} for related works in the case of the Friedrichs-Faddeev model.
\end{Remark}

\subsection{Finite rank perturbation}\label{finite_rank}

Here we apply the theory of Section \ref{Preliminaries} to finite rank perturbations of $H_0=Q$. Given
$u,v\in\H(\R)$ we write $P_{u,v}$ for the rank one operator $P_{u,v}:=\<u,\;\!\cdot\;\!\>v$, and we set
$P_v:=P_{v,v}$. The full Hamiltonian we consider is defined as follows.

\begin{Assumption}\label{assumption_H}
Fix an integer $N\ge0$ and take $\mu\ge0$. For $j,k\in\{1,\ldots,N\}$, let $v_j\in\H^\mu(\R)$ satisfy $\<v_j,v_k\>=\delta_{jk}$, and let $\lambda_j\in\R$. Then $H:=H_0+V$, where
$V:=\sum_{j=1}^N\lambda_jP_{v_j}$.
\end{Assumption}

\noindent
Many functions $v_j$ (as the Hermite functions \cite[p. 142]{RSI}) satisfy the requirements of
Assumption \ref{assumption_H}. Under Assumption \ref{assumption_H} the perturbation $V$ is bounded
from $\H^{-\mu}(\R)$ to $\H^\mu(\R)$, $H$ is selfadjoint on $\dom(H)=\dom(H_0)$, and the wave operators
$W_\pm$ exist and are complete \cite[Thm. XI.8]{RSIII}.

In the next lemma we establish some of the spectral properties of $H$, we prove a limiting
absorption principle for $H$, and we give a class of locally $H$-smooth operators. The limiting
absorption principle is expressed in terms of the Besov space
$\K:=(\H^1(\R),\H(\R))_{1/2,1}\equiv\H^{1/2,1}(\R)$ defined by real interpolation \cite[Sec. 3.4.1]{ABG}. We
recall that for each $s>1/2$ we have the continuous embeddings
$$
\H^s(\R)\subset\K\subset\H(\R)\subset\K^*\subset\H^{-s}(\R).
$$
We refer the reader to \cite[Sec. 6.2.1]{ABG} for the definition of the regularity classes
$C^k(A)$ and to \cite[Sec. 7.2.2]{ABG} for the definition of a (strict) Mourre estimate. The symbol
$\C_\pm$ stands for the half-plane $\C_\pm:=\{z\in\C\mid\pm\im(z)>0\}$.

\begin{Lemma}\label{rhube}
Let $H$ satisfy Assumption \ref{assumption_H} with $\mu\ge2$. Then
\begin{enumerate}
\item[(a)] $H$ has at most a finite number of eigenvalues, and each of these eigenvalues is of finite
multiplicity.
\item[(b)] The map $z\mapsto(H-z)^{-1}\in\B(\K,\K^*)$, which is holomorphic on $\C_\pm$, extends to
a weak* continuous function on $\C_\pm\cup\{\R\setminus\sigma_{\rm pp}(H)\}$. In particular, $H$ has
no singularly continuous spectrum.
\item[(c)] If $T$ belongs to $\B\big(\H^{-s}(\R),\H(\R)\big)$ for some $s>1/2$, then $T$ is locally $H$-smooth on
$\R\setminus\sigma_{\rm pp}(H)$.
\end{enumerate}
\end{Lemma}

The spectral results of points (a) and (b) on the finiteness of the singular spectrum of $H$ are not
surprising; they are known in the more general setting where $V$ is an integral operator with
H\"older continuous kernel (see \eg \cite[Thm. 1]{DNY91} and \cite[Lemma 3.10]{Fad64}). Note
however that point (a) implies that the sets $\D^s(\R)$ are dense in $\H(\R)$ for each $s\ge0$.

\begin{proof}
(a) Let $A:=-P$, then $\e^{-itA}H_0\e^{itA}=H_0+t$ for each $t\in\R$. Thus $H_0$ is of class
$C^\infty(A)$ and satisfies a strict Mourre estimate on $\R$ \cite[Sec. 7.6.1]{ABG}. Furthermore the
quadratic form
$$
\dom(A)\ni\varphi\mapsto\<\varphi,iVA\varphi\>-\<A\varphi,iV\varphi\>
$$
extends uniquely to the bounded form defined by the rank $2N$ operator
$F_1:=\sum_{j=1}^N\lambda_j\big(P_{v_j,v_j'}+P_{v_j',v_j}\big)$. This means that $V$ is of class
$C^1(A)$. Thus $H$ is of class $C^1(A)$ and since $F_1$ is compact, $H$ satisfies a Mourre estimate
on $\R$. The claim then follows by \cite[Cor. 7.2.11]{ABG}.

(b) The quadratic form
$$
\dom(A)\ni\varphi\mapsto\<\varphi,iF_1A\varphi\>-\<A\varphi,iF_1\varphi\>
$$
extends uniquely to the bounded form defined by the rank $3N$ operator
$F_2:=-\sum_{j=1}^N\lambda_j\big(P_{v_j'',v_j}+2P_{v_j',v_j'}+P_{v_j,v_j''}\big)$. This, together with
\cite[Thm. 7.2.9 \& Thm. 7.2.13]{ABG} and the proof of point (a), implies that $H$ is of class
$C^2(A)$ and that $H$ satisfies a strict Mourre estimate on $\R\setminus\sigma_{\rm pp}(H)$. It follows by
\cite[Thm. 01]{Sah97} (which applies to operators without spectral gap) that the map
$z\mapsto(H-z)^{-1}\in\B(\K,\K^*)$ extends to a weak* continuous function on
$\C_\pm\cup\{\R\setminus\sigma_{\rm pp}(H)\}$. In particular, $H$ has no singularly continuous
spectrum in $\R\setminus\sigma_{\rm pp}(H)$. Since continuous Borel measures on $\R$ have no pure
points \cite[p. 22]{RSI} and since $\sigma_{\rm pp}(H)$ is finite by point (a), we even get that $H$
has no singularly continuous spectrum at all.

(c) Since $T$ belongs to $\B\big(\dom(H),\H(\R)\big)$ and $T^*\H(\R)\subset\H^s(\R)\subset\K$, the claim is a consequence
of \cite[Prop. 7.1.3.(b)]{ABG} and the discussion that follows.
\end{proof}

We now study the differentiability of the function $x\mapsto S(x)$, which relies on the
differentiability of the boundary values of the resolvent of $H$.

\begin{Lemma}\label{higher_order}
Let $H$ satisfy Assumption \ref{assumption_H} with $\mu\ge n+1$ for some integer $n\ge1$. Let
$I\subset\{\R\setminus\sigma_{\rm pp}(H)\}$ be a relatively compact interval, and take $s>n-1/2$.
Then for each $x\in I$ the limits
$$
R^n(x\pm i0):=\lim_{\varepsilon\searrow0}(H-x\mp i\varepsilon)^{-n}
$$
exist in the norm topology of $\B\big(\H^s(\R),\H^{-s}(\R)\big)$ and are H\"older continuous. Furthermore
$x\mapsto R(x\pm i0)$ is $n-1$ times (H\"older continuously) differentiable as a map from $I$
to $\B\big(\H^s(\R),\H^{-s}(\R)\big)$, and
$$
\frac{\d^{n-1}}{\d x^{n-1}}\;\!R(x\pm i0)=(n-1)!\;\!R^n(x\pm i0).
$$
\end{Lemma}

\begin{proof}
The claims follow from \cite[Thm. 2.2.(iii)]{JMP84} applied to our
situation. We only have to verify the hypotheses of that theorem, namely that $H$ is $n$-smooth
with respect to $A=-P$ in the sense of \cite[Def. 2.1]{JMP84}. This is done in points (a), (b),
($\rm c_n$), ($\rm d_n$), and (e) that follow.

(a) $\dom(A)\cap\dom(H)\supset\S$ is a core for $H$.

(b) Let $\varphi\in\H_1(\R)$ and $\theta\in\R$. Then one has
$$
\|\e^{i\theta A}\varphi\|_{\H_1(\R)}
=\|\<Q+\theta\>\varphi\|
\le\big\|\<Q+\theta\>\<Q\>^{-1}\big\|\cdot\|\varphi\|_{\H_1(\R)}
\le2^{-1/2}(2+|\theta|)^{1/2}\|\varphi\|_{\H_1(\R)}.
$$
In particular, $\e^{i\theta A}$ maps $\dom(H)$ into $\dom(H)$, and
$\sup_{|\theta|\le1}\|H\e^{i\theta A}\varphi\|<\infty$ for each $\varphi\in\dom(H)$.

($\rm c_n$)-($\rm d_n$) Due to the proof Lemma \ref{rhube}.(a) the quadratic form
$$
\dom(A)\cap\dom(H)\ni\varphi\mapsto\<H\varphi,iA\varphi\>-\<A\varphi,iH\varphi\>
$$
extends uniquely to the bounded form defined by the operator $iB_1:=1+F_1$, where
$F_1=\sum_{j=1}^N\lambda_j\big(P_{v_j',v_j}+P_{v_j,v_j'}\big)$. Similarly for $j=2,3,\ldots,n+1$
the quadratic form
$$
\dom(A)\cap\dom(H)\ni\varphi\mapsto\<(iB_{j-1})^*\varphi,iA\varphi\>
-\<A\varphi,i(iB_{j-1})\varphi\>
$$
extends uniquely to a bounded form defined by an operator $iB_j:=F_j$, where $F_j$ is a
linear combination of the rank one operators $P_{v^{(j-k)},v^{(k)}}$, $k=0,1,\ldots,j$.

(e) Due to the proof Lemma \ref{rhube}.(a), $H$ satisfies a Mourre estimate on $\R$.
\end{proof}

For $m=1,2,\ldots,N$ let $V_m:=\sum_{j=1}^m\lambda_jP_{v_j}$ and $H_m:=H_0+V_m$. Then
it is known that the scattering matrix $S(x)$ factorizes for a.e. $x\in\R$ as
\cite[Eq. (8.4.2)]{Yaf92}
\begin{equation}\label{S_product}
S(x)=\widetilde{S_N}(x)\cdots\widetilde{S_2}(x)\widetilde{S_1}(x),
\end{equation}
where $\widetilde{S_m}(x)$ is unitarily equivalent to the scattering matrix $S_m(x)$
associated with the pair $\{H_m,H_{m-1}\}$. Since the difference $H_m- H_{m-1}$ is of rank one,
one can even obtain an explicit expression for $S_m(x)$ (see \cite[Eq. (6.7.9)]{Yaf92}).
For instance one has the following simple formula for $S_1(x)$ \cite[Eq. (8.4.1)]{Yaf92},
\cite[Eq. (66a)]{GP70}
$$
S_1(x)=\frac{1+\lambda_1F(x-i0)}{1+\lambda_1F(x+i0)}\;\!,
$$
where
$$
F(x\pm i0):=\lim_{\varepsilon\searrow0}\<v_1,(H_0-x\mp i\varepsilon)^{-1}v_1\>.
$$
Clearly Formula \eqref{S_product} is not very convenient for studying the differentiability of
the function $x\mapsto S(x)$. This is why we prove the usual formula for $S(x)$ in the next lemma.

Given $\tau\in\R$, we
let $\gamma(\tau):\S(\R)\to\C$ be the restriction operator defined by $\gamma(\tau)\varphi:=\varphi(\tau)$.
Some of the regularity properties of $\gamma(\tau)$ are collected in the appendix. Here we only recall that
$\gamma(\tau)$ extends uniquely to an element of $\B\big(\H^s(\R),\C\big)$ for each $s>1/2$.

\begin{Lemma}\label{representation}
Let $H$ satisfy Assumption \ref{assumption_H} with $\mu\ge2$. Then for each
$x\in\R\setminus\sigma_{\rm pp}(H)$ one has the equality
\begin{equation}\label{S_formula}
S(x)=1-2\pi i\gamma(x)[1-VR(x+i0)]V\gamma(x)^*.
\end{equation}
\end{Lemma}
\begin{proof}
The claim is a consequence of the stationary method for trace class perturbations \cite[Thm. 7.6.4]{Yaf92}
applied to the pair $\{H_0,H\}$.

The perturbation $V$ can be written as a product $V=G^*G_0$, with $G:=\sum_{j=1}^N\lambda_jP_{v_j}$ and
$G_0:=\sum_{j=1}^NP_{v_j}$. Since the operators $G$ and $G_0$ are selfadjoint and belong to the
Hilbert-Schmidt class, all the hypotheses of \cite[Thm. 7.6.4]{Yaf92} (and thus of
\cite[Thm. 5.7.1]{Yaf92}) are trivially satisfied. Therefore one has for a.e. $x\in\R$ the equality
\begin{equation}\label{Yafaev}
S(x)=1-2\pi i\gamma(x)G\big[1-\widetilde B(x+i0)\big]G_0\gamma(x)^*,
\end{equation}
where $\widetilde B(x+i0)$ is the norm limit defined by the condition
$$
\lim_{\varepsilon\searrow0}\big\|G_0(H-x-i\varepsilon)^{-1}G-\widetilde B(x+i0)\big\|=0.
$$
On another hand we know from Lemma \ref{higher_order} that the limit $R(x+i0)$ exists in the
norm topology of $\B\big(\H^s(\R),\H^{-s}(\R)\big)$ for each $x\in\R\setminus\sigma_{\rm pp}(H)$ and each
$s>1/2$. Since we also have $G_0,G\in\B\big(\H^{-\mu}(\R),\H^\mu(\R)\big)$, we get the identity
$\widetilde B(x+i0)=G_0R(x+i0)G$. This together with Formula \eqref{Yafaev} implies the claim.
\end{proof}

We are in a position to show the differentiability of the scattering matrix.

\begin{Lemma}\label{S_differentiable}
Let $H$ satisfy Assumption \ref{assumption_H} with $\mu\ge n+1$ for some integer $n\ge1$. Then
$x\mapsto S(x)$ is $n-1$ times (H\"older continuously) differentiable from $\R\setminus\sigma_{\rm pp}(H)$
to $\C$.
\end{Lemma}

\begin{proof}
Due to Formula \eqref{S_formula} it is sufficient to prove that the terms
$$
A(x):=\big(\textstyle\frac{\d^{\ell_1}}{\d x^{\ell_1}}\;\!\gamma(x)\big)V
\big(\textstyle\frac{\d^{\ell_2}}{\d x^{\ell_2}}\;\!\gamma(x)^*\big)
$$
and
$$
B(x):=\big(\textstyle\frac{\d^{\ell_1}}{\d x^{\ell_1}}\;\!\gamma(x)\big)V
\big(\textstyle\frac{\d^{\ell_2}}{\d x^{\ell_2}}\;\!R(x+i0)\big)V
\big(\textstyle\frac{\d^{\ell_3}}{\d x^{\ell_3}}\;\!\gamma(x)^*\big)
$$
exist and are locally H\"older continuous on $\R\setminus\sigma_{\rm pp}(H)$ for all non-negative
integers $\ell_1,\ell_2,\ell_3$ satisfying $\ell_1+\ell_2+\ell_3\le n-1$. The factors in $B(x)$
satisfy
\begin{align*}
\big(\textstyle\frac{\d^{\ell_3}}{\d x^{\ell_3}}\;\!\gamma(x)^*\big)
&\in\B\big(\C,\H^{-s_3}(\R)\big)\quad{\rm for}~s_3>\ell_3+1/2,\\
V&\in\B\big(\H^{-s_3}(\R),\H^{s_2}(\R)\big)\quad{\rm for}~s_2,s_3\in[0,\mu],\\
\big(\textstyle\frac{\d^{\ell_2}}{\d x^{\ell_2}}\;\!R(x+i0)\big)
&\in\B\big(\H^{s_2}(\R),\H^{-s_2}(\R)\big)\quad{\rm for}~s_2>\ell_2+1/2,\\
V&\in\B\big(\H^{-s_2}(\R),\H^{s_1}(\R)\big)\quad{\rm for}~s_1,s_2\in[0,\mu],\\
\big(\textstyle\frac{\d^{\ell_1}}{\d x^{\ell_1}}\;\!\gamma(x)\big)
&\in\B\big(\H^{s_1}(\R),\C\big)\quad{\rm for}~s_1>\ell_1+1/2,
\end{align*}
and are locally H\"older continuous due to Lemma \ref{higher_order} and Lemma \ref{gamma_smooth}.
Therefore if the $s_j$'s above are chosen so that $s_j\in(\ell_j+1/2,\mu]$ for $j=1,2,3$, then
$B(x)$ is finite and locally H\"older continuous on $\R\setminus\sigma_{\rm pp}(H)$. Since similar
arguments apply to the term $A(x)$, the claim is proved.
\end{proof}

\begin{Lemma}\label{lone_conditions}
Let $H$ satisfy Assumption \ref{assumption_H} with $\mu>2$. Then one has for each $\varphi\in\H^s(\R)$,
$s>2$,
\begin{equation}
\left\|(W_--1)\e^{-itH_0}\varphi\right\|\in\lone(\R_-,\d t)\label{R-}
\end{equation}
and
\begin{equation}
\left\|(W_+-1)\e^{-itH_0}\varphi\right\|\in\lone(\R_+,\d t).\label{R+}
\end{equation}
\end{Lemma}

\begin{proof}
For $\varphi\in\H^s(\R)$ and $t\in\R$, we have (see \eg the proof of \cite[Lemma 4.6]{Jen81})
$$
\(W_--1\)\e^{-itH_0}\varphi
=-i\e^{-itH}\int_{-\infty}^t\d\tau\,\e^{i\tau H}V\e^{-i\tau H_0}\varphi,
$$
where the integral is strongly convergent. Hence to prove \eqref{R-} it is enough to show that
\begin{equation}\label{l1 condition}
\int_{-\infty}^{-\delta}\d t\int_{-\infty}^t\d\tau\left\|V\e^{-i\tau H_0}\varphi\right\|<\infty
\end{equation}
for some $\delta>0$. Let $\zeta:=\min\{\mu,s\}$, then $\big\|\<P\>^\zeta\varphi\big\|$ and
$\big\|V\<P\>^\zeta\big\|$ are finite by hypothesis. If $|\tau|$ is big enough, it follows by
\eqref{evolution} that
\begin{align*}
\big\|V\e^{-i\tau H_0}\varphi\big\|
\le{\rm Const.}\;\!\big\|\<P\>^{-\zeta}\e^{-i\tau Q}\<P\>^{-\zeta}\big\|
&={\rm Const.}\;\!\big\|\<P-\tau\>^{-\zeta}\<P\>^{-\zeta}\big\|\\
&\le{\rm Const.}\;\!|\tau|^{-\zeta}.
\end{align*}
Since $\zeta>2$, this implies \eqref{l1 condition}, and thus \eqref{R-}. The proof of \eqref{R+}
is similar.
\end{proof}

In the next theorem we prove Formula \eqref{Eisenbud} for Hamiltonians $H$ satisfying Assumption \ref{assumption_H} with $\mu\ge5$.

\begin{Theorem}\label{final_one}
Let $f\ge0$ satisfy Assumption \ref{assumption_Frie}, and let $H$ satisfy Assumption
\ref{assumption_H} with $\mu\ge5$. Then one has for each $\varphi\in\D^3(\R)$ the identity
$$
\lim_{r\to\infty}\tau_r^{\rm in}(\varphi)
=\lim_{r\to\infty}\tau_r(\varphi)
=-i\int_\R\d x\,|\varphi(x)|^2\overline{S(x)}S'(x).
$$
\end{Theorem}

\begin{proof}
Let $\varphi\in\D^3(\R)$. Then $S\varphi\in\D^3(\R)$ by Lemma \ref{S_differentiable}, and conditions
\eqref{H-}-\eqref{H+} are verified by Lemma \ref{lone_conditions}. Therefore all the hypotheses of
Theorem \ref{abs_delay} and Remark \ref{Remark_Eisenbud} are satisfied, and so the claim is
proved.
\end{proof}

\section*{Acknowledgements}

The author thanks the Swiss National Science Foundation for financial support. This
work was completed while the author was visiting the University of Chile. He would
like to thank Professor M. M\u antoiu for his kind hospitality.

\section*{Appendix}

We collect in this appendix some facts on the restriction operator $\gamma(\tau)$ of Lemma
\ref{representation}. We consider the general case with configurations space $\R^d$, $d\ge1$.

Given $\tau\in\R$, we let $\gamma(\tau):\S(\R^d)\to\ltwo(\R^{d-1})$ be the restriction operator defined by $\gamma(\tau)\varphi:=\varphi(\tau,\cdot)$. We know from \cite[Thm. 2.4.2]{Kur78} that
$\gamma(\tau)$ extends uniquely to an element of $\B\big(\H^s(\R^d),\ltwo(\R^{d-1})\big)$ for each $s>1/2$.
Furthermore $\gamma(\tau)$ is H\"older continuous in $\tau$ with respect to the operator norm,
namely for all $\tau,\tau'\in\R$ there exists a constant $\textsc c$ such that
\begin{equation}\label{Holder}
\big\|\gamma(\tau)-\gamma(\tau')\big\|_{\B(\H^s(\R^d),\ltwo(\R^{d-1}))}\le\textsc c
\begin{cases}
|\tau-\tau'|^{s-1/2} & \textrm{if }s\in\big(\12,\frac32\big),\\
|\tau-\tau'|\cdot|\ln|\tau-\tau'|| & \textrm{if }s=\frac32\textrm{ and }|\tau-\tau'|<\12,\\
|\tau-\tau'| & \textrm{if }s>\frac32.
\end{cases}
\end{equation}
Finally $\gamma(\tau)$ has the following differentiability property.

\begin{Lemma}\label{gamma_smooth}
Let $s>k+\12$ with $k\ge0$ integer. Then $\gamma$ is $k$ times (H\"older continuously)
differentiable as a map from $\R$ to $\B\big(\H^s(\R^d),\ltwo(\R^{d-1})\big)$.
\end{Lemma}

\begin{proof}
We adapt the proof of \cite[Lemma 3.3]{Jen81}. Consider first $s>k+\12$ with $k=1$. The
obvious guess for the derivative at $\tau$ of $\gamma$ is
$({\sf D}\gamma)(\tau):=\gamma(\tau)\partial_1$, where $\partial_1$ stands for the partial
derivative w.r.t. the first variable. Thus one has for $\varphi\in\S(\R^d)$ and
$\delta\in\R$ with $|\delta|\in(0,1/2)$
$$
\big\{{\textstyle\frac1\delta}[\gamma(\tau+\delta)-\gamma(\tau)]-({\sf D}\gamma)(\tau)\big\}\varphi
={\textstyle\frac1\delta}\int_0^\delta\d\xi
\big[(\partial_1\varphi)(\tau+\xi,\;\!\cdot\;\!)-(\partial_1\varphi)(\tau,\;\!\cdot\;\!)\big].
$$
In particular, using the first (and thus the most pessimistic) bound in \eqref{Holder}, we get
\begin{align*}
&\big\|\big\{{\textstyle\frac1\delta}[\gamma(\tau+\delta)
-\gamma(\tau)]-({\sf D}\gamma)(\tau)\big\}\varphi\big\|_{\ltwo(\R^{d-1})}\\
&\le{\textstyle\frac1{|\delta|}}\int_0^{|\delta|}\d\xi\;\!
\big\|(\partial_1\varphi)(\tau+\sgn(\delta)\xi,\;\!\cdot\;\!)
-(\partial_1\varphi)(\tau,\;\!\cdot\;\!)\big\|_{\ltwo(\R^{d-1})}\\
&\le\|\partial_1\varphi\|_{\H^{s-1}(\R^d)}\;\!{\textstyle\frac1{|\delta|}}\int_0^{|\delta|}\d\xi\,
\|\gamma(\tau+\sgn(\delta)\xi)-\gamma(\tau)\|_{\B(\H^{s-1}(\R^d),\ltwo(\R^{d-1}))}\\
&\le{\rm Const.}\,\|\varphi\|_{\H^s(\R^d)}\;\!{\textstyle\frac1{|\delta|}}\int_0^{|\delta|}\d\xi\,
|\xi|^{s-3/2}\\
&\le{\rm Const.}\,\|\varphi\|_{\H^s(\R^d)}|\delta|^{s-3/2}.
\end{align*}
Since $\S(\R^d)$ is dense in $\H^s(\R^d)$ and ${\sf D}\gamma:\R\to\B\big(\H^s(\R^d),\ltwo(\R^{d-1})\big)$
is H\"older continuous, this proves the result for $k=1$. The result for $k>1$ follows then easily
by using the expression for $({\sf D}\gamma)(\tau)$.
\end{proof}


\def\cprime{$'$}

\end{document}